\documentclass[usenatbib]{mn2e}

\usepackage[dvips]{graphicx}
\usepackage{amssymb}
\usepackage{txfonts}

\newcommand{\xte}{{\textit{RXTE}}}

\newcommand{\sax}{{\textit{Beppo\-SAX}}}

\newcommand{\fermi}{{\textit{Fermi}}}
\newcommand{\agile}{{\textit{AGILE}}}
\newcommand{\swift}{{\textit{Swift}}}
\newcommand{\msun}{{\rm M}_{\sun}}
\newcommand{\rsun}{{\rm R}_{\sun}}
\newcommand{\g}{$\gamma$}

\let\oldhat\hat
\renewcommand{\vec}[1]{\mathbf{#1}}
\renewcommand{\hat}[1]{\oldhat{\mathbf{#1}}}

\newbox\grsign \setbox\grsign=\hbox{$>$} \newdimen\grdimen \grdimen=\ht\grsign
\newbox\simlessbox \newbox\simgreatbox \newbox\simpropbox
\setbox\simgreatbox=\hbox{\raise.5ex\hbox{$>$}\llap
     {\lower.5ex\hbox{$\sim$}}}\ht1=\grdimen\dp1=0pt
\setbox\simlessbox=\hbox{\raise.5ex\hbox{$<$}\llap
     {\lower.5ex\hbox{$\sim$}}}\ht2=\grdimen\dp2=0pt
\setbox\simpropbox=\hbox{\raise.5ex\hbox{$\propto$}\llap
     {\lower.5ex\hbox{$\sim$}}}\ht2=\grdimen\dp2=0pt

\title[The gamma-ray emitting jet of Cyg X-3]{The gamma-ray emitting region of the jet in Cyg X-3}

\author[A. A. Zdziarski et al.]
{Andrzej A. Zdziarski,$^{1}$\thanks{E-mail: aaz@camk.edu.pl, sikora@camk.edu.pl} Marek Sikora,$^1$\footnotemark[1] Guillaume Dubus,$^2$ Feng Yuan,$^3$  
\newauthor Benoit Cerutti$^4$ and Anna Ogorza{\l}ek$^5$\\
$^1$Centrum Astronomiczne im.\ M. Kopernika, Bartycka 18, PL-00-716 Warszawa, Poland\\
$^2$UJF-Grenoble 1/CNRS-INSU, Institut de Plan{\'e}tologie et d'Astrophysique de Grenoble, UMR 5274, 38041 Grenoble, France\\
$^3$Key Laboratory for Research in Galaxies and Cosmology, Shanghai Astronomical Observatory, Chinese Academy of Sciences,80 Nandan Road, Shanghai\\ 200030, China\\
$^4$Center for Integrated Plasma Studies, Physics Department, University of Colorado, Boulder, CO 80309, USA\\
$^5$ Astronomical Observatory, Jagiellonian University, Orla 171, 30-244
Krak\'ow, Poland
}

\date{Accepted 2012 January 9.  Received 2012 January 7; in original form 2011 November 3}

\pagerange{\pageref{firstpage}--\pageref{lastpage}}
\pubyear{2012}

\begin{document}

\maketitle

\label{firstpage}

\begin{abstract}
We study models of the \g-ray emission of Cyg X-3 observed by \fermi. We calculate the average X-ray spectrum during the \g-ray active periods. Then, we calculate spectra from Compton scattering of a photon beam into a given direction by isotropic relativistic electrons with a power-law distribution, both based on the Klein-Nishina cross section and in the Thomson limit. Applying the results to scattering of stellar blackbody radiation in the inner jet of Cyg X-3, we find that a low-energy break in the electron distribution at a Lorentz factor of $\sim 300$--$10^3$ is required by the shape of the observed X-ray/\g-ray spectrum in order to avoid overproducing the observed X-ray flux. The electrons giving rise to the observed \g-rays are efficiently cooled by Compton scattering, and the power-law index of the acceleration process is $\simeq 2.5$--3. The bulk Lorentz factor of the jet and the kinetic power before the dissipation region depend on the fraction of the dissipation power supplied to the electrons; if it is $\simeq 1/2$, the Lorentz factor is $\sim 2.5$, and the kinetic power is $\sim 10^{38}$ erg s$^{-1}$, which represents a firm lower limit on the jet power, and is comparable to the bolometric luminosity of Cyg X-3. Most of the power supplied to the electrons is radiated. The broad band spectrum constrains the synchrotron and self-Compton emission from the \g-ray emitting electrons, which requires the magnetic field to be relatively weak, with the magnetic energy density $\la$ a few times $10^{-3}$ of that in the electrons. The actual value of the magnetic field strength can be inferred from a future simultaneous measurement of the IR and \g-ray fluxes.  
\end{abstract}

\section{Introduction}
\label{intro}

Cyg X-3 is a high-mass X-ray binary with a Wolf-Rayet (WR) companion \citep{v96}, with an unusually short orbital period of 4.8 h, located at a distance $D\simeq\! 7$--9 kpc in the Galactic plane \citep*{lzt09,d83,p00}. Due to the lack of reliable mass functions and determination of the inclination, the nature of its compact object remains uncertain (see \citealt{v09} for a recent discussion). However, the presence of a black hole is favoured by considering the X-ray and radio emission and the bolometric luminosity \citep*{sz08,szm08,h08,h09}. Also, \citet*{zmg10} have shown that the differences in the form of the X-ray spectra of Cyg X-3 from those of confirmed black-hole binaries can be accounted for by Compton scattering in the very strong stellar wind from the companion. That model also accounts for the lack of high frequencies \citep*{alh09} in the power spectra of Cyg X-3. 

\begin{figure*}
\centerline{\includegraphics[width=18cm]{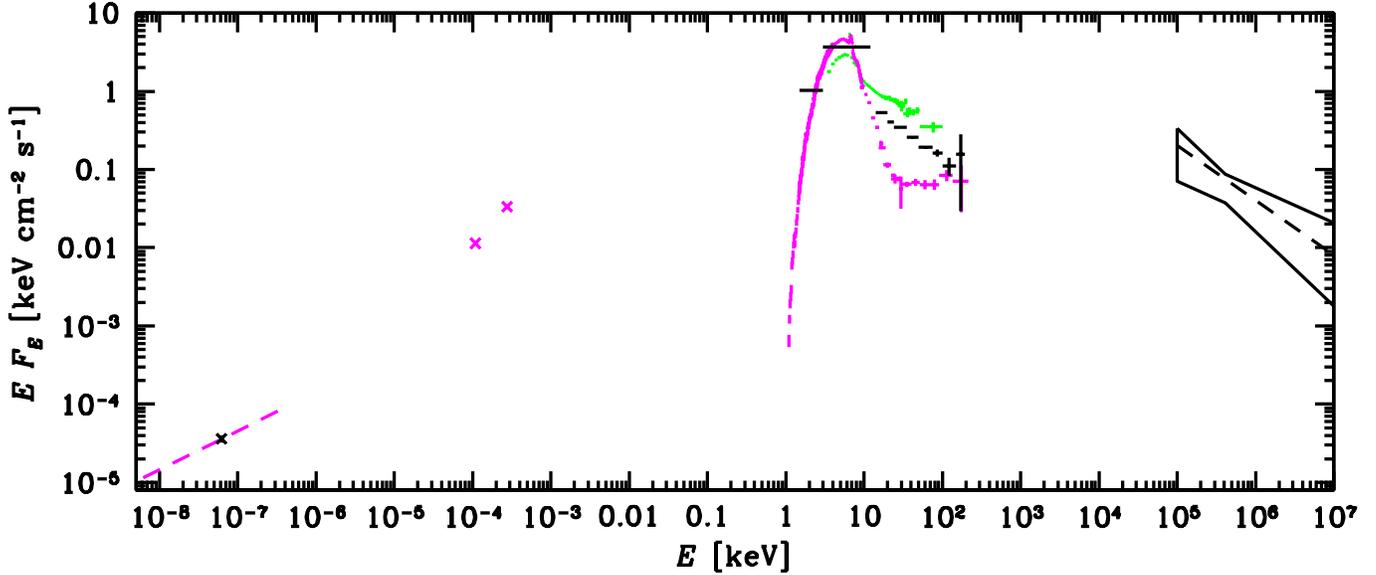}} 
\caption{The average radio to \g-ray spectrum of Cyg X-3 during the 2008 and 2009 \g-ray active periods is shown in the black symbols. The error contour and the black dashed line show the average power-law fit with $\Gamma=2.70\pm 0.25$ to the \fermi/LAT data (FLC09). The black error bars show the simultaneous X-ray spectrum from the \xte/ASM and \swift/BAT. The black cross shows the average 15 GHz radio flux measured by the AMI and OVRO telescopes during the same periods. The magenta and green error bars show two other soft-state X-ray spectra, from \sax\/ and \xte, respectively. The magenta dashed line shows a 1.5--100 GHz, $F_\nu\propto \nu^{-0.5}$, spectrum, similar to the spectra measured in that range during a 2001 radio outburst of Cyg X-3, and the two magenta crosses show IR measurements in a radio-flaring state. See Section \ref{data} for details.
} \label{r_x_gamma}
\end{figure*}

Cyg X-3 is a persistent X-ray source. Its X-ray spectra have been classified into five states by \citet{szm08}, who have also quantified their correlations with the radio states. Its \g-ray emission has been discovered by the \fermi\/ Large Area Telescope (LAT) and \agile\/ in the soft spectral states (\citealt{fermi}, hereafter FLC09; \citealt{agile}). Fig.\ \ref{r_x_gamma} shows the average power-law \g-ray emission measured by \fermi\/ during the active phases. The power law is relatively steep, with the photon index of $\Gamma\simeq 2.70\pm 0.25$. Fig.\ \ref{r_x_gamma} also shows the average X-ray spectrum and the 15 GHz flux during the periods of \g-ray emission. It also shows two X-ray spectra and the IR fluxes in the soft state from earlier observations (see Section \ref{data}).

We see that the \g-ray power law spectrum has to have a low-energy break somewhere between $\sim$1 and $\sim$100 MeV in order not to produce more X-ray emission than observed. Furthermore, \citet{z12} show that the hard X-rays up to at least 100 keV during the intervals with \g-ray emission have the orbital modulation pattern characteristic to wind absorption and scattering. On the other hand, the GeV orbital modulation is shifted in phase with respect to the X-rays by $\sim\upi/2$ (FLC09). This appears to imply that the contribution of the spectral component observed in the GeV range to the 100 keV flux is at most weak. The GeV power law emission and its orbital modulation appear to be due to Compton up-scattering of the stellar emission from the companion WR star by relativistic electrons in the jet of this source (FLC09; \citealt*{dch10b}, hereafter DCH10). 

In this work, we first calculate the average X-ray and radio emission during the \g-ray active periods. Then, we study emission due to Compton up-scattering of blackbody photons by power-law electrons with a low-energy break. Finally, we apply our theoretical results to the broad-band X-ray/\g-ray (X\g) spectra of Cyg X-3, and obtain strong constraints on the electron distribution in the \g-ray emitting region and on the parameters of the jet.

\section{The radio--X-ray spectra}
\label{data}

We use X-ray monitoring data from the \swift\/ Burst Alert Telescope (BAT; \citealt{barthelmy05,m05}) in the form of a 14--195 keV 8-channel light curve \citep{z12}. The channels are between energies of 14, 20, 24, 35, 50, 75, 100, 150 and 195 keV. Also, we use X-ray monitoring data from the the All-Sky Monitor (ASM; \citealt*{brs93,levine96}) on board {\it Rossi X-ray Timing Explorer\/} (\xte). The ASM has three channels at energies of 1.5--3 keV, 3--5 keV and 5--12 keV. In the radio range, we use the 15 GHz data from the AMI Large Array and the Owens Valley Radio Observatory (OVRO), which data were used in FLC09. 

We then calculate the average fluxes in each of the X-ray channels (converting the count rates to fluxes using the method of \citealt*{zps11}) and at 15 GHz during the two \g-ray active periods for which the average \g-ray spectrum of FLC09 was obtained, which are MJD 54750--54820 (in 2008) and MJD 54990--55045 (in 2009). The resulting average X-ray spectrum and the average radio flux simultaneous with the \g-ray spectrum are shown in Fig.\ \ref{r_x_gamma}. The average 15 GHz flux equals $0.38\pm 0.04$ Jy, with a large rms of 0.43 Jy, reflecting a strong flux variability from a few mJy to $\sim$2 Jy. 

To illustrate the likely form of the radio spectrum during the \g-ray active periods, we plot a radio power law between 1.5 GHz and 100 GHz with an energy index of 0.5 (characteristic to uncooled optically-thin synchrotron emission), which is in the middle of the $\sim$0.4--0.6 index range measured in that frequency range during a 2001 radio outburst of Cyg X-3 by \citet{mj04}. To show the likely form of the broad-band soft-state radio-flaring spectrum of Cyg X-3, we also show two IR measurements, at 4.5 $\mu$m and 11.5 $\mu$m, taken in the flaring radio state on 1997 June 18, with the quiescent-state fluxes (which are, most likely, due to the stellar wind emission) subtracted \citep{ogley01}. The 15 GHz flux measured by the Ryle telescope at the time of the IR measurement, MJD 506217.(39--42),  was 0.64 Jy, i.e., about 1.5 the average value during the \g-ray active periods, and the 8.3 GHz flux from the Green Bank Interferometer was 0.59--0.72 Jy. (Note that the MJD of the measurements given in \citealt{ogley01} are in error.)

In Fig.\ \ref{r_x_gamma}, we also show one of the average soft-state X-ray spectra from \xte\/ Proportional Counter Array (PCA) and the High Energy X-Ray Transient Experiment (HEXTE) of \citet{szm08}, namely their spectrum \#4, and the soft spectrum from \sax\/ of \citet{sz08}, which is similar, but of better quality, to the spectrum \#5 of \citet{szm08}. We see that our average ASM/BAT spectrum lies between these two spectra, thus it is intermediate between the soft and ultrasoft states of \citet{szm08}. As shown by \citet{zmg10}, there is a close correspondence between the canonical X-ray states of black-hole binaries and the X-ray states of Cyg X-3. The soft and ultrasoft states in black-hole binaries are dominated by emission of an optically-thick accretion disc up to an energy of several keV, and by a power-law like tail, probably of coronal origin, at higher energies. The same situation is then, most likely, present in those states of Cyg X-3, with a modification due to the passing of that emission through the very strong stellar wind of the donor WR star. Some jet contribution is also possible at hard X-rays, which issue is studied in this work.

\section{Anisotropic Compton scattering}
\label{compton}

In this Section, we consider Compton scattering of soft photons by a cloud of relativistic electrons isotropic in the jet comoving frame. Thus, the photon energy, electron Lorentz factor and the scattering angle below are given in this frame. 

The problem of Compton scattering of a mono-directional photon beam by a cloud of relativistic electrons with a Lorentz factor $\gamma\gg 1$ and an isotropic angular distribution into a given angle has been solved by \citet{aa81}. Their equation (20), valid from the Thomson to the Klein-Nishina regimes, gives the flux per electron and per solid angle, which can be written as,
\begin{eqnarray}
\lefteqn{{\epsilon{\rm d} \dot n(\epsilon_0,\gamma)\over {\rm d}\epsilon{\rm d}\Omega}= {3\sigma_{\rm T} \dot n_0 \epsilon\over 16\upi\epsilon_0\gamma^2} \left[\left(1-r\right)^2+r^2+{w^2\over 2(1-w)}
\right] \,\,{\rm s}^{-1},\label{rate}
}\\
\lefteqn{ r\equiv {\epsilon\over 2\epsilon_0 x\gamma^2 (1-w)}\leq 1,\quad x\equiv 1-\cos \vartheta,\quad w\equiv {\epsilon\over \gamma}, \label{def}}
\end{eqnarray}
where $\epsilon_0$ and $\epsilon$ are the energy of the incoming and scattered photon, respectively, in units of $m_{\rm e}c^2$, $\vartheta$ is the scattering angle, $\dot n_0$ [cm$^{-2}$ s$^{-1}$] is the number flux of incoming photons, $m_{\rm e}$ is the electron mass, and $\sigma_{\rm T}$ is the Thomson cross section. Also,
\begin{equation}
\epsilon_0\ll \epsilon\leq {2 x\epsilon_0\gamma^2\over 1+2 x\epsilon_0\gamma},
\label{range}
\end{equation}
where the former constraint expresses the applicability of the $\gamma^2\gg 1$ condition, and the latter, equivalent to $r\leq 1$, is kinematic. 

\begin{figure}
\centerline{\includegraphics[width=\columnwidth]{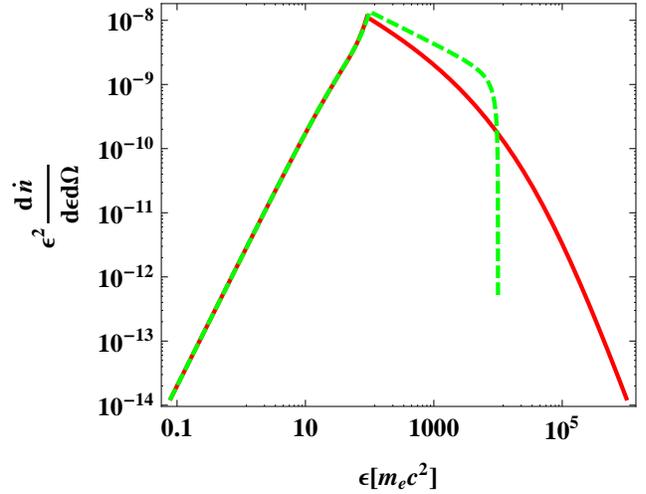}} 
\caption{An example spectrum from a cloud of electrons irradiated by a beam of monoenergetic photons with $\epsilon_0=5\times 10^{-5}$, $\dot n_0=1$, emitted at $\vartheta=90\degr$ with respect to the beam for $p=4$, $\gamma_1=10^3$, $\gamma_2\rightarrow \infty$, and the normalization corresponding to $K=1$. The red solid and green dashed curves correspond to the Klein-Nishina formula (\ref{pl_rate}) and the Thomson limit with a sharp Klein-Nishina cut-off, equations (\ref{rate0}--\ref{power}), respectively.
} \label{mono}
\end{figure}

The above rate can be then integrated over an electron distribution. We consider the case of a power-law distribution with cut-offs, 
\begin{equation}
N(\gamma) =\cases{K \gamma^{-p}, &$\gamma_1\leq\gamma\leq \gamma_2$;\cr
0, &otherwise,\cr}
\label{ngamma}
\end{equation}
where the constant $K$ specifies either the electron density or their total number. Accounting for $r\leq 1$, we have,
\begin{equation}
{\epsilon {\rm d} \dot n(\epsilon_0)\over {\rm d}\epsilon{\rm d}\Omega}=\int\limits_{\min\left\{\max\left[\gamma_1, {\epsilon\over 2} \left(1+\sqrt{ 1+2/(\epsilon_0\epsilon x)}\right)\right],\gamma_2\right\} }^{\gamma_2}
{\epsilon {\rm d} \dot n(\epsilon_0,\gamma)\over {\rm d}\epsilon{\rm d}\Omega} N(\gamma) {\rm d}\gamma.
\label{integration}
\end{equation}
For $\gamma_2\rightarrow \infty$, this yields,
\begin{eqnarray}
\lefteqn{\nonumber {\epsilon {\rm d} \dot n(\epsilon_0)\over {\rm d}\epsilon{\rm d}\Omega}={3\sigma_{\rm T} \dot n_0 K\over 32\upi\epsilon^{2+p}\epsilon_0^3 x^2}
\left\{ y^{1+p}\left[{2(\epsilon\epsilon_0 x)^2\over 1+p}+{(4+p)y^2\over 3+p}+ {y^3\over 1-y}\right]+
\right.} \\
\lefteqn{\label{pl_rate} \left. \left[(\epsilon\epsilon_0 x-1)^2 -5-p\right] {\rm B}_y (3+p,0)\right\},}\\
\lefteqn{y=\cases{\displaystyle{\epsilon/ \gamma_1}, &$\displaystyle{\epsilon_0\geq {\epsilon\over 2\gamma_1 x(\gamma_1 -\epsilon)}\,{\rm and}\,\epsilon<\gamma_1}$;\cr
\displaystyle{2/\left(1+\sqrt{1+{2\over \epsilon\epsilon_0 x}}\right),} &otherwise,\cr} \label{y}}
\end{eqnarray}
where ${\rm B}_y$ is the incomplete beta function. In terms of $\epsilon$, the first condition in equation (\ref{y}) reads $\epsilon\leq 2 x\epsilon_0\gamma_1^2/( 1+2 x\epsilon_0 \gamma_1)$. Equation (\ref{pl_rate}) for the second case in equation (\ref{y}) is equivalent to equation (33) of \citet{aa81}. For integer or half-integer $s$, ${\rm B}_y(s,0)$ can be expressed relatively simply by elementary functions, and, in general,
\begin{equation}
{\rm B}_y(s,0)=\sum_{j=0}^\infty {y^{s+j}\over s+j}, \quad y<1.
\label{beta}
\end{equation}
For a finite $\gamma_2$, the flux can be obtained by subtracting the rate with $\gamma_2$ substituted for $\gamma_1$ in equation (\ref{y}) from the rate of equation (\ref{pl_rate}). Then, the spectrum will be null for $\epsilon\geq 2 x\epsilon_0\gamma_2^2/( 1+2 x\epsilon_0\gamma_2)$. Fig.\ \ref{mono} shows an example of the spectrum for parameters roughly applicable to Cyg X-3, and compares it to the Thomson-limit spectrum of Appendix \ref{thomson}. We see that for the chosen parameters, the average slope of the actual spectrum for a decade above the break (corresponding to $\gamma_1$) is substantially steeper than of that in the Thomson limit. 

\begin{figure}
\centerline{\includegraphics[width=\columnwidth]{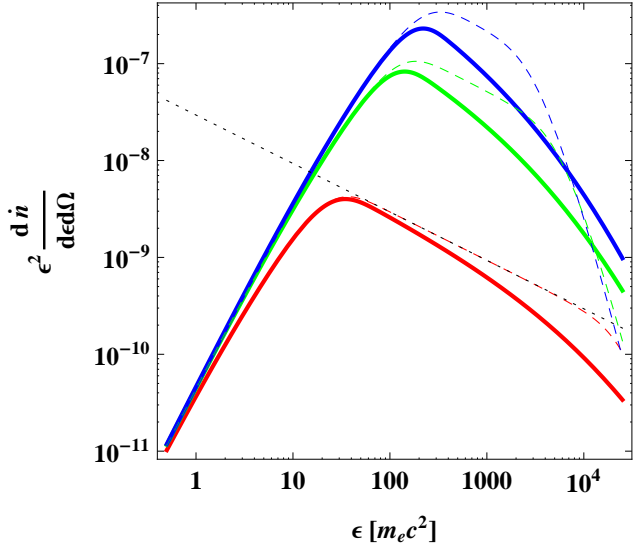}} 
\caption{The thick solid blue, green and red curves show example Klein-Nishina spectra in the cloud frame from a stationary ($\beta_{\rm j}=0$) cloud of electrons irradiated by a beam of blackbody photons with the emission at an angle with respect to the beam given by $\cos\vartheta=-0.8$, 0, and 0.8, respectively, for $p=4$, $\gamma_1=10^3$, $\gamma_2\rightarrow \infty$, $T=10^5$ K, $K=1$, $R=R_*$. The corresponding dashed curves show the Thomson-limit spectra, see Appendix \ref{thomson}. The black dotted line shows the spectrum of equation (\ref{bb_rate}) for $\cos\vartheta=0.8$.
} \label{beam}
\end{figure}

Equation (\ref{pl_rate}) can be then integrated piece-wise over a distribution of irradiating photons, $\dot n_0(\epsilon_0)$, 
\begin{equation}
{\epsilon {\rm d} \dot n\over {\rm d}\epsilon{\rm d}\Omega}= \int_{\epsilon/ [2 x \gamma_2^2 (1-\epsilon/\gamma_2)]}^{q\epsilon} {\dot n_0(\epsilon_0)\over \dot n_0} {\epsilon {\rm d} \dot n(\epsilon_0)\over {\rm d}\epsilon{\rm d}\Omega} {\rm d}\epsilon_0,
\label{int_rate}
\end{equation}
where $q\ll 1$ is a constant assuring the $\epsilon_0\ll \epsilon$ condition, at which equation (\ref{rate}) is valid. When $K$ corresponds to electron density, ${\epsilon {\rm d} \dot n/ {\rm d}\epsilon{\rm d}\Omega}$ is the emissivity. When $K$ corresponds to the total number of electrons, it gives the total photon production rate, and the dimension of equation (\ref{int_rate}) is s$^{-1}$. In the latter case, the differential luminosity and the observed flux (neglecting relativistic corrections) are given by,
\begin{equation}
{{\rm d} L\over {\rm d}\epsilon{\rm d}\Omega}= m_{\rm e} c^2 {\epsilon {\rm d} \dot n\over {\rm d}\epsilon{\rm d}\Omega}, \quad {{\rm d} F\over {\rm d}\epsilon}= {1\over D^2} {{\rm d} L\over {\rm d}\epsilon{\rm d}\Omega},
\label{LF}
\end{equation}
respectively. 

The seed photons can be, in particular, from a blackbody emitter (e.g., a star) with a radius, $R_*$, and temperature, $T$. When the distance of the electron cloud from the stellar centre, $R$, is $\gg R_*$, the blackbody photons form an almost mono-directional beam, incident on the electrons. Here, we take into account that the electron cloud may be located in a jet or counterjet moving with respect to the star, for which the Doppler factor is,
\begin{equation}
{\cal D}_{*}={1\over \Gamma_{\rm j}(1-\beta_{\rm j}\, \vec{e}_* \!\cdot\! \vec{e}_{\rm j})},
\label{dstar}
\end{equation}
where $\beta_{\rm j}$ is the jet velocity, $\Gamma_{\rm j}=1/(1-\beta_{\rm j}^2)^{1/2}$ is the jet Lorentz factor, and $\vec{e}_*$ and $\vec{e}_{\rm j}$, are the unit vectors along the direction from the star towards the electron cloud, and along the jet, respectively, see fig.\ 1 in DCH10. Then, the soft photon energy in the stellar (= observer) frame is $\epsilon_0 {\cal D}_*$, and this energy has a blackbody distribution at $k T/m_{\rm e}c^2$. Since $\dot n_0(\epsilon_0)$ is a relativistic invariant \citep{bg70},
\begin{equation}
\dot n_0(\epsilon_0)={2\upi\over c^2 h^3}\left(R_*\over R\right)^2 {(m_{\rm e}c^2)^3\epsilon_0^2 {\cal D}_*^2\over \exp(\epsilon_0 m_{\rm e}c^2 {\cal D}_*/k T)-1}\,\, {\rm cm^{-2}\, s^{-1}},
\label{bb}
\end{equation}
where $k$ and $h$ are the Boltzmann and Planck constants, respectively.

Fig.\ \ref{beam} shows example spectra obtained using equation (\ref{int_rate}) for blackbody irradiation, also comparing it to the corresponding Thomson-limit spectra of Appendix \ref{thomson}. Such a model can give the low-energy break required by the broad-band X\g\ spectrum observed from Cyg X-3, see Fig.\ \ref{r_x_gamma}. We see that the photon break energy moves to lower energies as $\vartheta$ decreases, as implied by equation (\ref{y}). The low-energy parts of the spectra are almost independent of the angle. Similarly to the case of mono-energetic incident photons, Fig.\ \ref{mono}, the Thomson-limit spectra above the break have the slope significantly harder than the actual spectra, but the difference between the two decreases with the decreasing scattering angle, $\vartheta$. 

\section{Observed spectra}
\label{observed}

Here, we calculate the spectra in the observer's frame. We consider a steady-state jet, in which the observed emission comes from a given spatial range in the observer's frame. This is compatible with the dynamical time scale of the jet of the order of tens of seconds (see Section \ref{parameters}) whereas the \g-ray emission was detected over time scales of days/weeks. Furthermore, the strong orbital modulation in \g-rays (FLC09) requires the \g-ray emitting region to be approximately stationary. 

The jet and counterjet Doppler factors with respect to the observer are
\begin{equation}
{\cal D}_{\rm j}={1\over \Gamma_{\rm j}(1-\beta_{\rm j} \vec{e}_{\rm obs} \!\cdot\! \vec{e}_{\rm j})},\quad {\cal D}_{\rm cj}={1\over \Gamma_{\rm j}(1+\beta_{\rm j} \vec{e}_{\rm obs} \!\cdot\! \vec{e}_{\rm j})},
\label{dobs}
\end{equation}
respectively, where $\vec{e}_{\rm obs}$ is the unit vector towards the observer. We use the result of \citet*{dch10a},
\begin{equation}
x\equiv 1-\cos\vartheta = {\cal D}_{\rm j} {\cal D}_* \left(1-\vec{e}_{\rm obs} \!\cdot\! \vec{e}_*\right),
\label{x}
\end{equation}
where $\vec{e}_{\rm obs} \!\cdot\! \vec{e}_*$ is the cosine of the orbital-phase dependent angle between the direction from the star to the jet and from the jet to the observer in the observer's frame (which angle in the jet frame is the scattering angle, $\vartheta$). For the jet emission,
\begin{equation}
\epsilon={E\over {\cal D}_{\rm j} m_{\rm e} c^2},\quad
{{\rm d}F\over {\rm d}E }={{\cal D}_{\rm j}^2\over D^2 \Gamma_{\rm j}} {\epsilon {\rm d} \dot n\over {\rm d}\epsilon {\rm d}\Omega},
\label{ef}
\end{equation}
where $E$ is the observed dimensional photon energy, and the flux transformation to the observed frame is for a steady-state jet \citep{s97}. We note that the form of ${\rm d}F/ {\rm d}E$ above assumes the the energy units in $F$ and $E$ are the same, whereas they are often assumed to be different (e.g, erg and eV), which, however, can be easily accounted for. For the counterjet, ${\cal D}_{\rm cj}$ and the corresponding ${\cal D}_*$, $R$ should be used, and the two observed fluxes should be added. Given the observed spectrum, this transformation yields the normalization of the emitting electron distribution corresponding to the actual number of electrons in the considered jet region. On the other hand, the transformation used in DCH10, with ${\cal D}_{\rm j}^3$ instead of ${\cal D}_{\rm j}^2/\Gamma_{\rm j}$, corresponds either to emission of a single moving blob or to the observed number of electrons in the emitting part of the jet. Note that in the latter case the observed number is different in the jet and the counterjet, which was not accounted for in the treatment used by DCH10.

\begin{figure}
\centerline{\includegraphics[width=\columnwidth]{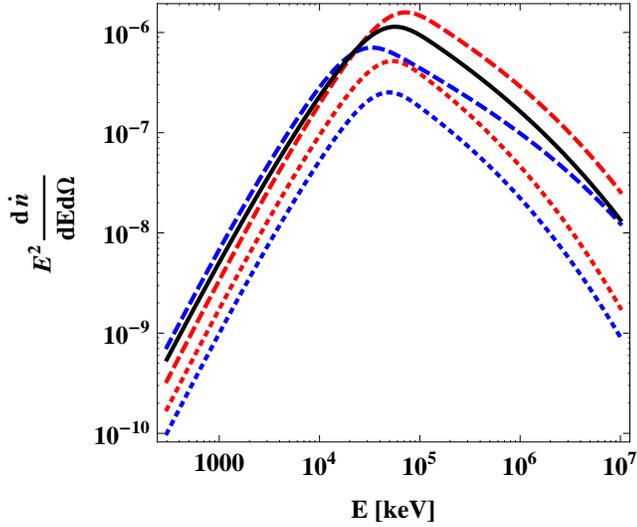}} 
\caption{Example spectra for the best-fit black-hole model of DCH10 in the observer's frame, but using the full Klein-Nishina spectra, $p=4$, and modified to account for the flux transformation of equation (\ref{ef}), see Section \ref{observed}. The dashed red and blue curves give the spectra from the jet at the orbital phases (defined as in DCH10) of $\theta=0.2\upi$ and $0.7\upi$, respectively, and the corresponding dotted curves give the spectra from the counterjet. The solid black curve gives the spectrum from both the jet and counterjet averaged over the orbit. The parameters are $\gamma_1=10^3$, $\gamma_2\rightarrow \infty$, $T=10^5$ K. The normalization corresponds to $K=1$; then the unit of vertical axis is keV s$^{-1}$.
} \label{g0_1000}
\end{figure}

For a pure power-law emission from Thomson-limit scattering, as in equation (\ref{bb_rate}), applying the above relations results in the energy flux as given by equations (1) and (3) of DCH10 (except for the form of the dependence on ${\cal D}_{\rm j}$), with no dependence on ${\cal D}_*$. In a general case, we need to apply formulae (\ref{x}--\ref{ef}) to equation (\ref{int_rate}) (i.e., substitute $\epsilon$ and $x$ as above and multiply the rate by ${\cal D}_{\rm j}^2/\Gamma_{\rm j}$), integrate it numerically, and then repeat the procedure for the counterjet.

The distance between the electron cloud and the stellar centre, $R$, and the components of the vectors $\vec{e}_*$, $\vec{e}_{\rm obs}$ and $\vec{e}_{\rm j}$ are given by equations (4) and (5) of DCH10, respectively. They depend on several parameters of the system, namely, the orbital separation, $d$, the height of the electron cloud along the jet, $H$, the binary inclination, $i$, the inclination of the jet with respect to the normal to the binary plane, given by the azimuth, $\theta_{\rm j}$ and the polar angle, $\phi_{\rm j}$, and the orbital phase, $\theta$ (note that DCH10 use a non-standard definition of $\theta$, with $\theta=\pm \upi/2$ rather than the usual 0, $\upi$, at the conjunctions). We include here both the jet and the counterjet, for which the polar angle is $\phi_{\rm j}+\upi$ (and its unit vector is $-\vec{e}_{\rm j}$). We note that the model of DCH10 neglects eclipses of the counterjet by the star. We neglect them here as well for consistency with the adopted assumption that both the blackbody emission and the scattered emission are point-like.

We use here the model of DCH10 with their assumed black-hole binary parameters, with $d= 4.1\times 10^{11}$ cm, $i=30\degr$, $T=10^5$ K and $R_*=2.3\rsun$. However, since we use a different relativistic transformation between the jet and observer's frames, we have refitted the model to the observed orbital modulation of \g-rays using equation (\ref{ef}). We have obtained $H=7.6\times 10^{11}$ cm, $\theta_{\rm j}=319\degr$, $\phi_{\rm j}=39\degr$, which are the same as in DCH10, but $\beta_{\rm j}=0.47$ ($\Gamma_{\rm j}\simeq 1.13$), which is somewhat higher than their value of 0.41.

At these parameters, relativistic beaming is moderate, ${\cal D}_{\rm j}\simeq 1.50$, ${\cal D}_{\rm cj}\simeq 0.62$, ${\cal D}_*\simeq 1.45$--1.60, $1-\vec{e}_{\rm obs} \!\cdot\! \vec{e}_*\simeq 0$--0.4 for the jet and 1.6--2 for the counterjet, and $R/R_*\simeq 3.6$--6.7. Since we use here the Klein-Nishina cross section instead of the Thomson approximation, the actual spectrum is softer than in the Thomson limit, where $p=2\Gamma-1=4.4$. Here, we use $p=4$, which approximately reproduces the observational best-fit spectrum. 

Fig.\ \ref{g0_1000} shows two example spectra for the electron distribution with $\gamma_1=10^3$ and $\theta=0.2\upi$ and $0.7\upi$. We see they look similar to those in Fig.\ \ref{beam}, except for an additional shift of their relative normalization, introduced by the phase-dependent ${\cal D}_*$ and $R/R_*$. We then average the observed spectrum over the orbital phase, $\theta$, which we plot in the solid curve. 

\begin{figure*}
\centerline{\includegraphics[width=18cm]{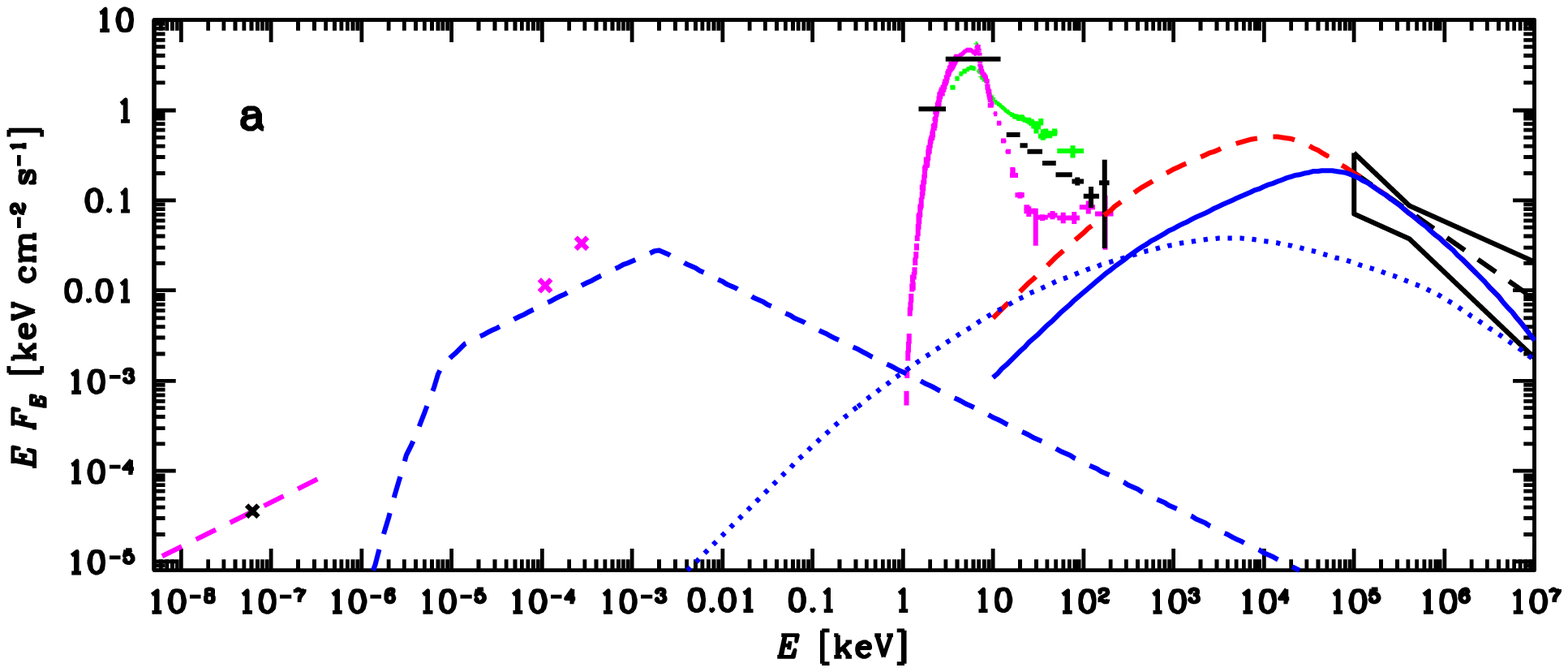}} 
\centerline{\includegraphics[width=18cm]{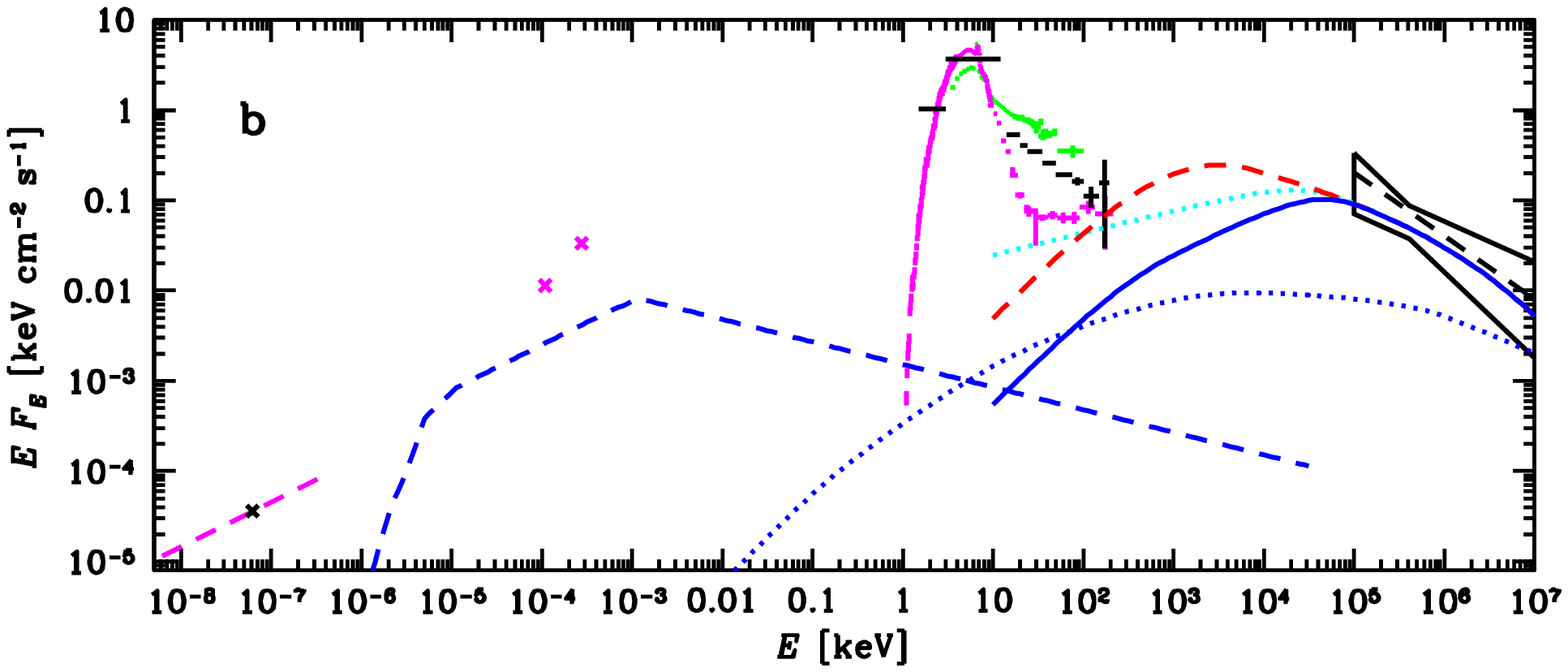}} 
\caption{The radio to \g-ray spectra spectra (the same as in Fig.\ \ref{r_x_gamma}) compared to the Compton and synchrotron models for the electron distribution of equation (\ref{ngamma1}) with $\gamma_{\rm b}=130$, $\gamma_2\rightarrow \infty$, $T=10^5$ K. (a) The red dashed and blue solid curves show the Compton spectra for $\gamma_1=700$ and 1500, respectively, and $p=4$. The model spectra are normalized to the best-fit average \fermi\/ spectrum. The dashed blue curve shows the synchrotron spectrum from the \g-ray producing electrons for $\gamma_1=1500$ and the magnetic field given by $\eta_B=10^{-3}$. The blue dotted curve shows the synchrotron self-Compton spectrum. (b) The same for $p=3.5$ and $\gamma_1=300$ and 1300, for the red dashed and blue solid, dashed, and dotted curves, respectively. The normalization of the solid curves is below the best fit but within the error contour. We see that the value of the electron low-energy break is constrained to $300\la \gamma_1\la 10^3$, but models with $\gamma_1< 10^3$ give strong contribution to the hard X-rays. The cyan dotted curve shows the Compton spectrum corresponding to the electron distribution of equation (\ref{ngamma2}) with $\gamma_{\rm b}=10^3$ and $\gamma_1= 10$. This model is in principle possible, but the value of $\gamma_{\rm b}$ used for this model is several times our estimate, equation (\ref{gb2}).
} \label{comparison}
\end{figure*}

We note, however, that we have to take into account the electrons below $\gamma_1$. Even if the electrons are accelerated only above $\gamma_1$, they lose energy via Compton, synchrotron, and adiabatic losses and form a distribution below $\gamma_1$. Hereafter a dot will denote a time derivative in the jet frame (as in Section \ref{compton}), and d.../d$t$ will denote a time derivative in the observer's frame. The loss rates in the jet frame are given by
\begin{eqnarray}
\lefteqn{
\dot\gamma_{\rm C}={4 f_{\rm KN}\sigma_{\rm T} U_{\rm rad}\gamma^2\over 3m_{\rm e} c}, \quad U_{\rm rad}= {2\upi^5(kT)^4 \over 15 c^3 h^3} \left(R_*\over R {\cal D}_*\right)^2,\label{gdot1}}\\
\lefteqn{
\dot\gamma_{\rm S}={\sigma_{\rm T} B^2\gamma^2\over 6\upi m_{\rm e} c}, \quad \dot\gamma_{\rm ad}\simeq {2\beta_{\rm j} \Gamma_{\rm j} \gamma c\over 3 H},
\label{gdot2}}
\end{eqnarray}
respectively, and $U_{\rm rad}$ is the blackbody energy density within the electron cloud, $f_{\rm KN}<1$ gives the Klein-Nishina reduction with respect to the Thomson limit, $B$ is the magnetic field strength, and the factor of $2/3$ in $\dot\gamma_{\rm ad}$ accounts for the expansion being in two dimensions only. The cooling rate for a single electron for monoenergetic seed photons using the Klein-Nishina cross section was calculated by \citet{jones65,jones68}. The Lorentz factor at which the Compton and adiabatic rates equal each other, and the electron distribution has a break, is
\begin{equation}
\gamma_{\rm b}={15\over 4\upi^5}{m_{\rm e} c^5 h^3\over \sigma_{\rm T}f_{\rm KN} (kT)^4}{\beta_{\rm j}\Gamma_{\rm j}\over H} \left(R{\cal D}_*\over R_*\right)^2,
\label{gb1}
\end{equation}
which for the assumed parameters equals to,
\begin{equation}
\gamma_{\rm b}\simeq 130 \left(T\over 10^5\,{\rm K}\right)^{-4} {0.018\over (R_*/R {\cal D}_*)^2},
\label{gb2}
\end{equation}
where $f_{\rm KN}=1$ was assumed and the factor of 0.018 is the value of the orbital overage of $(R_*/R {\cal D}_*)^2$ in $U_{\rm rad}$. Thus, the break Lorentz factor at our parameters is well below the minimum Lorentz factor of $\sim 10^3$ required to explain the observed \g-ray spectrum. We note that the dependence of $\gamma_{\rm b}$ on $H$ and $\beta_{\rm j}$ is rather complex; $H^{-1}$ appears in equation (\ref{gb1}), but $R$ also depends on $H$ and ${\cal D}_*$ depends on $H$ and $\beta_{\rm j}$. Furthermore, the parameters are mutually connected via the requirement of fitting the observed orbital modulation.

We assume that the electrons are accelerated at a power-law rate, which in either the jet or counterjet frame is given by,
\begin{equation}
Q(\gamma)\simeq K_{\rm inj} \gamma^{1-p},\quad \gamma_1\leq \gamma\leq \gamma_2. 
\label{q_gamma}
\end{equation}
where $K_{\rm inj}$ is the normalization factor. Hereafter, $K$ and $K_{\rm inj}$ correspond to the total number of electrons in the jet (and not to their density). Then, assuming Compton losses in the Thomson limit ($f_{\rm KN}=1$) and synchrotron losses, and for $\gamma_1>\gamma_{\rm b}$ (fast cooling), the steady-state distribution (in either jet or counterjet) will approximately be,
\begin{equation}
N(\gamma) \simeq \cases{K \gamma_1^{2-p}\gamma_{\rm b}^{-1} \gamma^{-1}, &$\gamma_0\leq\gamma\leq \gamma_{\rm b}$;\cr
K \gamma_1^{2-p}\gamma^{-2}, &$\gamma_{\rm b}\leq\gamma\leq \gamma_1$;\cr
K \gamma^{-p}, &$\gamma_1\leq\gamma\leq \gamma_2$;\cr
0, &otherwise,\cr}
\label{ngamma1}
\end{equation}
where $\gamma_0\sim 1$ is a minimum overall Lorentz factor. The steady-state electron kinetic equation in the comoving frame is $N(\gamma)=\dot \gamma^{-1} \int_\gamma^{\infty} Q(\gamma){\rm d}\gamma$, where $\dot \gamma$ is total loss rate. Assuming the dominant losses above $\gamma_1$ are Compton in the Thomson regime, $K_{\rm inj}$ is then related to $K$ by,
\begin{equation}
K_{\rm inj} = {4 \sigma_{\rm T} U_{\rm rad}(p-2)K\over 3 m_{\rm e}c}.
\label{Kinj}
\end{equation}

Fig.\ \ref{comparison}(a) compares the models with $p=4$ with the observations. Here, we impose $K$ to match the best-fit \fermi\/ spectrum and require that both the low-energy break is at $\la 0.1$ GeV, and the X-rays at $\sim$100 keV are not overproduced. We show the average spectra for the electron distribution of equation (\ref{ngamma1}) with $\gamma_{\rm b}=130$ and $\gamma_1=700$ and 1500. As expected, the low-energy break is at an energy $\propto \gamma_1^2$. Fig.\ \ref{comparison}(b) compares the models with $p=3.5$ and the normalization somewhat below the best fit, which both are within the observational uncertainties. We see that these results imply a constraint of $300\la \gamma_1\la 1500$. However, models with $\gamma_1< 10^3$ give strong contribution to the hard X-rays, which appears in conflict with the related result that the orbital modulation at $\sim 100$ keV during the \g-ray emitting intervals is characteristic to bound-free absorption and Compton scattering by the stellar wind and out of phase with the $>0.1$ GeV modulation \citep{z12}. We note that the index of the accelerated electrons in our models is $p-1\simeq 2.5$--3. 

\begin{figure}
\centerline{\includegraphics[width=\columnwidth]{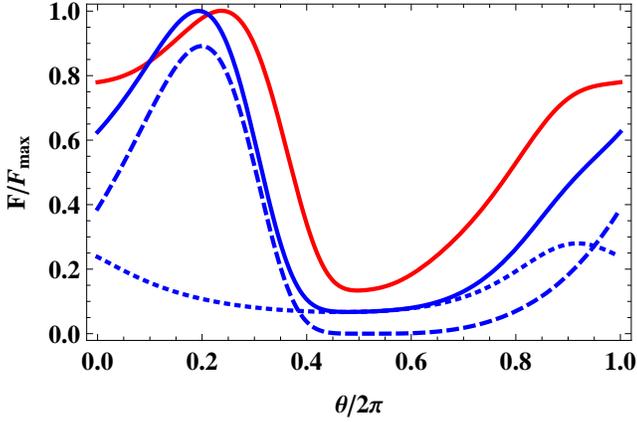}} 
\caption{The orbital modulation of the (fast-cooling) model with $\gamma_1=700$, $\gamma_{\rm b}=130$, and $p=4$ (which spectrum is shown by the red dashed curve in Fig.\ \ref{comparison}a) at $10^2$ keV (red solid curve) and $0.2$ GeV (blue solid curve). The blue dashed and dotted curves show separately the contributions of the jet and counterjet at 0.2 GeV. The modulation pattern at $10^2$ keV of the (slow-cooling) model shown by the cyan dotted curve in Fig.\ \ref{comparison}(b) is relatively similar to that shown by the blue solid curve. The superior conjunction (compact object behind the WR star) is at $\theta/2\upi =0.25$, around which the X-rays show the minimum due the maximum absorption/scattering. On the other hand, our models have the maxima around this phase, which appears to rule out any substantial contribution of the jet to the X-rays.
} \label{orb_mod}
\end{figure}

The red and blue curves in Fig.\ \ref{orb_mod} show the orbital modulation pattern at 100 keV and 0.2 GeV of the model shown Fig.\ \ref{comparison}(a) with $p=4$ and $\gamma_1=700$ (at 0.2 GeV, the modulation pattern of the models with $\gamma_1=700$ and $1500$ are identical). We see that the patterns at the two energies are very similar, and have the maxima around the superior conjunction, whereas the X-rays have the minimum at this phase. Given that the optical depth through the wind from the electron cloud is much lower than that from around the compact object, we can use the orbital modulation of X-rays to distinguish the X-ray source location close to the compact object from that in the scattering cloud in the jet.

On the other hand, the parameters used in equation (\ref{gb2}), e.g., $T$, $R_*$, bear large uncertainties, and we cannot exclude in principle a much larger value of $\gamma_{\rm b}$. For $\gamma_1<\gamma_{\rm b}$ (slow cooling), we would then have approximately, 
\begin{equation}
N(\gamma) \simeq \cases{K \gamma_{\rm b}^{-1}\gamma_1^{2-p}\gamma^{-1}, &$\gamma_0\leq\gamma\leq \gamma_1$;\cr
K \gamma_{\rm b}^{-1}\gamma^{1-p}, &$\gamma_1\leq\gamma\leq \gamma_{\rm b}$;\cr
K \gamma^{-p}, &$\gamma_{\rm b}\leq\gamma\leq \gamma_2$;\cr
0, &otherwise,\cr}
\label{ngamma2}
\end{equation}
The cyan curve in Fig.\ \ref{comparison}(b) shows the case for $p=3.5$, $\gamma_{\rm b}=10^3$, $\gamma_1= 10$. We consider this model relatively unlikely given the estimate of $\gamma_{\rm b}$ of equation (\ref{gb2}), and consider below models with $\gamma_{\rm b}< \gamma_1$. Although it reproduces well the high-energy tail of the softest X-ray spectrum, its orbital modulation is very similar to those shown in Fig.\ \ref{orb_mod}, which is in conflict with the observed hard X-ray modulation \citep{z12}. 

\section{The jet structure}
\label{jet}

\subsection{The jet power}
\label{j_power}

The amplitude of the \g-ray orbital modulation close to unity indicates the \g-ray emission region is rather compact, and the jet does not emit along a range of heights large compared to the average distance of the source from the compact object, $H$. Otherwise regions at different heights would have different modulation patterns, strongly reducing the net modulation amplitude. Thus, prior to the \g-ray emission, we assume non-radiating electron-ion jet and counterjet moving with the Lorentz factor of $\Gamma_{\rm j,0}$. The sum power of the jet and counterjet is then dominated by the bulk motion of cold ions, 
\begin{equation}
P_{\rm j,0}=2 m_{\rm i} c^2 (\Gamma_{\rm j,0}-1) {{\rm d}N_{\rm i}\over {\rm d}t},
\label{pj0}
\end{equation}
where ${\rm d}N_{\rm i}/{\rm d}t$ is the ion number flux in either the jet or counterjet in the observer's frame and $m_{\rm i}$ is the ion mass. Given that Cyg X-3 contains an He donor \citep*{v96,fhp99}, $m_{\rm i}\simeq 4 m_{\rm p}$, where $m_{\rm p}$ is the proton mass. The jet then enters a shock region, reducing its bulk Lorentz factor to $\Gamma_{\rm j}$, which for the best-fit parameters of the orbital modulation model is $\Gamma_{\rm j} \simeq 1.13$ (Section \ref{observed}). The fraction of the initial energy dissipated is,
\begin{equation}
\eta_{\rm diss}={\Gamma_{\rm j,0}-\Gamma_{\rm j}\over \Gamma_{\rm j,0}-1}.
\label{eta_diss}
\end{equation}
The power supplied to the electrons is given by,
\begin{equation}
P_{\rm e,inj}=\eta_{\rm e}\eta_{\rm diss}P_{\rm j,0}= 2 m_{\rm e}c^2\int Q(\gamma) \gamma{\rm d}\gamma.
\label{P_e}
\end{equation}
where $\eta_{\rm e}$ is the fraction of the dissipated energy supplied to electrons in the shock acceleration region. We assume (e.g., \citealt{s08}) that all electrons in the dissipative zone are accelerated/heated to relativistic energies. Thus,
\begin{equation}
{{\rm d}N_{\rm i}\over {\rm d}t}={n_{\rm i} \over n_{\rm e}} {{\rm d}N_{\rm e}\over {\rm d}t},\quad {{\rm d}N_{\rm e}\over {\rm d}t}={1\over \Gamma_{\rm j}} \int Q(\gamma) {\rm d}\gamma,
\label{ndoti}
\end{equation}
where ${\rm d}N_{\rm e}/{\rm d}t$ is the total electron number flux in the observer's frame, $Q(\gamma)$ is the electron acceleration/heating rate in the jet frame, and the ion/electron density ratio is $n_{\rm i}/n_{\rm e}\simeq 1/2$ for He in the absence of pair production. 

Combining equations (\ref{pj0}--\ref{ndoti}), we find,
\begin{equation}
\Gamma_{\rm j,0}=\Gamma_{\rm j}\left(1+{m_{\rm e}n_{\rm e}\over m_{\rm i} n_{\rm i}}{\bar\gamma_{\rm inj}\over \eta_{\rm e}}\right),
\label{dGamma}
\end{equation}
where $\bar \gamma_{\rm inj}$ is the average Lorentz factor of the accelerated/heated electrons in the jet frame. Here, we approximate $Q(\gamma)$, consisting of a relativistic Maxwellian and a power-law tail (e.g., \citealt{s08}), as a power law with the index of $p-1$ and a low energy cut-off, see equation (\ref{q_gamma}). Then, for $\gamma_2\rightarrow \infty$,
\begin{equation}
\bar\gamma_{\rm inj}={\int Q(\gamma)\gamma{\rm d}\gamma\over \int Q(\gamma){\rm d}\gamma}\simeq {\gamma_1 (p-2)\over p-3},\quad {{\rm d}N_{\rm e}\over {\rm d}t}={\gamma_1^{2-p} K_{\rm inj}\over \Gamma_{\rm j}(p-2)},
\label{gamma_av}
\end{equation}
for $p>3$ and $p>2$, respectively, and where $K_{\rm inj}$ is given by equation (\ref{Kinj}). Equation (\ref{dGamma}) then yields, for $p=4$, $\Gamma_{\rm j,0}\simeq \Gamma_{\rm j}+(0.62/ \eta_{\rm e})(\gamma_1/ 10^3)$. The value of $\eta_{\rm e}$ remains unknown; if the dissipated power is divided equally among electrons and ions, $\eta_{\rm e}\sim 1/2$. For this value and $\gamma_1=10^3$, $\Gamma_{\rm j,0}\simeq 2.38$ and $\eta_{\rm diss}\simeq 0.90$.

The radiated power, $P_{\rm rad}$, is related to the power in the electrons by,
\begin{equation}
P_{\rm rad}=\eta_{\rm rad}P_{\rm e,inj},
\label{prad}
\end{equation}
where $\eta_{\rm rad}$ is the radiation efficiency. We then calculate the initial jet power and the radiative power as
\begin{eqnarray}
\lefteqn{
P_{\rm j,0}=\left[{\Gamma_{\rm j}-1\over \Gamma_{\rm j}} {m_{\rm i}n_{\rm i}\over m_{\rm e}n_{\rm e}}+ {\gamma_1(p-2)\over (p-3)\eta_{\rm e}}\right]{8\gamma_1^{2-p} K\sigma_{\rm T} c U_{\rm rad}\over 3},
\label{pj0s}}\\
\lefteqn{
P_{\rm rad}=\eta_{\rm rad} {\gamma_1^{3-p}(p-2)\over p-3}{8K\sigma_{\rm T} c U_{\rm rad}\over 3}.
\label{prads}}
\end{eqnarray}
Note that in the above derivation we did not need to specify the extend of the dissipation zone. Since $\gamma_{\rm b}\ll \gamma_1$, see equation (\ref{gb2}), the radiative efficiency in the dissipation zone is $\eta_{\rm rad}\simeq 1$. 

The normalization constant in Fig.\ \ref{comparison}(a), for $p=4$, is $K/D^2= 2.1\times 10^5$, which implies $K\simeq 1.0\times 10^{50}(D/7\,{\rm kpc})^2$. (For the models in Fig.\ \ref{comparison}b with $p=3.5$, $K/D^2= 2.2\times 10^3$.) This yields the powers averaged over the orbital phase of $P_{\rm j,0}\simeq 7.2\left[0.11+(0.5/\eta_{\rm e})(\gamma_1/10^3)\right] (\gamma_1/10^3)^{2-p} (D/7\,{\rm kpc})^2 10^{37}$ erg s$^{-1}$, and $P_{\rm rad}\simeq 3.6 (\gamma_1/10^3)^{3-p} (D/7\,{\rm kpc})^2 10^{37}$ erg s$^{-1}$. (We note that the jet radiative output of electrons with $\gamma>10^3$ given in DCH10 is mistakenly too large by a factor of $4\upi$.) The Klein-Nishina corrected radiative power for the steady-state electron distribution of equation (\ref{ngamma1}) is slightly lower.

We can compare the total mass flow rate in the pre-shock jet, i.e., including the rest mass and the associated kinetic energy, $\dot M_{\rm j}=2 m_{\rm i} \Gamma_{\rm j,0} {\rm d}N_{\rm i}/{\rm d}t$, to the mass accretion rate estimated from the bolometric luminosity of the source. At $\eta_{\rm e}=0.5$, $\dot M_{\rm j}\simeq 1.5(D/7\,{\rm kpc})^2 10^{17}$ g s$^{-1}$. On the other hand, the bolometric luminosity of Cyg X-3 in the soft state calculated by \citet{sz08} is $\simeq 2(D/7\,{\rm kpc})^2 10^{38}$ erg s$^{-1}$, which, at an accretion efficiency of $\epsilon_{\rm accr}=0.1$, requires $\dot M_{\rm accr}\simeq 2(D/7\,{\rm kpc})^2 10^{18}$ g s$^{-1}$. Thus, a relatively small fraction of the mass flow rate at the outer boundary of the accretion source in its soft state is sufficient to power the jet, unless $\eta_{\rm e}\ll 1$. We note that this fraction is still much higher than that estimated for the hard-state jet in the black-hole binary XTE J1118--480 of $\sim 0.01$ by \citet*{ycn05}. The Eddington limit on the mass accretion rate for He corresponds to $\dot M_{\rm E}\simeq 3\times 10^{19}(M/10\msun) (\epsilon_{\rm accr}/0.1)^{-1}$ erg s$^{-1}$, where $M$ is the black-hole mass. Adopting this limit imposes a constraint on $\eta_{\rm e}$. 

\subsection{Electrons, cooling and pair production}
\label{parameters}

Based on the steady-state electron distribution of equation (\ref{ngamma1}), we can calculate the total number of relativistic electrons and their total energy in the comoving frame for the distribution of equation (\ref{ngamma1}) (for $\gamma_2\rightarrow\infty$) for either the jet and counterjet,
\begin{eqnarray}
\lefteqn{
N_{\rm e}=K \gamma_1^{1-p}\left[{\gamma_1\over \gamma_{\rm b}}\left(1+\ln{\gamma_{\rm b}\over \gamma_0}\right)-{p-2\over p-1}\right],\quad p>1,}\\
\lefteqn{
E_{\rm e}\simeq K m_{\rm e} c^2 \gamma_1^{2-p}\left({p-1\over p-2} +\ln{\gamma_1\over \gamma_{\rm b}}\right),\quad p>2, }
\end{eqnarray}
respectively. This, for $p\simeq 4$, $\gamma_{\rm b}=130$, $\gamma_1=10^3$, yields $N_{\rm e}\simeq 4.4 (D/7\,{\rm kpc})^2 10^{42}$ and $E_{\rm e} \simeq 2.1(D/7\,{\rm kpc})^2 10^{38} $ erg. Then, the radiative time scale averaged over the electron distribution in the jet frame is
\begin{equation}
\langle t_{\rm rad}\rangle \equiv { E_{\rm e}\over P_{\rm rad}/2}\simeq 11 \left(\gamma_1\over 10^3\right)^{-1}\left(T\over 10^5\,{\rm                                                     K}\right)^{-4}{\rm s}.
\label{trad}
\end{equation}
The cooling time of an individual electron, $\sim (10^4/\gamma)$ s [neglecting $f_{\rm KN}$, see equation (\ref{gdot1})], is approximately equal to $\langle t_{\rm rad}\rangle$ at $\gamma=\gamma_1$, which is close to the average electron energy, see equation (\ref{gamma_av}). Also, $\beta_{\rm j} \Gamma_{\rm j}c \langle t_{\rm rad}\rangle$ also gives the minimum size of the emission region. It cannot be more compact because the electrons would not then have time to cool.

The dynamical time scale, in the jet frame, is longer (which also follows from $\gamma_{\rm b}<\gamma_1$ for our parameters),
\begin{equation}
t_{\rm dyn}\equiv {H\over \beta_{\rm j}\Gamma_{\rm j} c}\simeq 50 \left(H\over 8\times 10^{11}\,{\rm cm}\right)\,{\rm s}.
\label{tdyn}
\end{equation}
Since the emission at $>0.1$ GeV is due to electrons with $\gamma>10^3$, their cooling time being $<t_{\rm dyn}$ is compatible with the orbital modulation of photons with energies $>0.1$ GeV being close to 100 per cent, requiring the corresponding emitting region to be compact. On the other hand, low-energy electrons will radiate and lose their energy over longer ranges of the jet length, and thus the depth of the orbital modulation is expected to decrease somewhat with the decreasing photon energy at $E< 0.1$ GeV.

\begin{figure}
\centerline{\includegraphics[width=6.5cm]{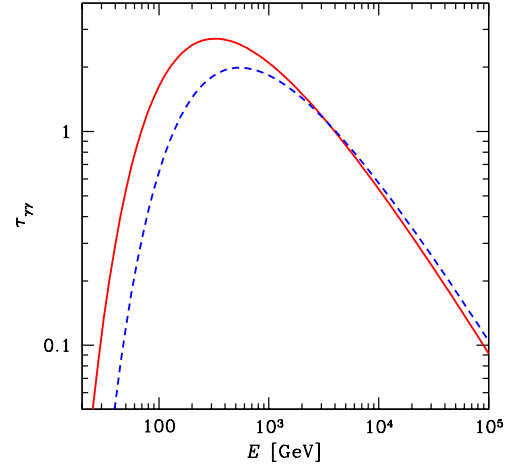}} 
\caption{The optical depth to pair production on stellar photons for \g-rays produced in the electron cloud. The red solid and blue dashed curves are for $\theta=0.9$ and 2.0 (in radians), respectively. The optical depth is around the maximum for the former phase, and it goes to a minimum of $\tau_{\gamma\gamma} \ll 1$ at all energies for $\theta\simeq 3.2$. The phase of $\theta=2.0$ is intermediate. The distance from the stellar centre and the angle to the observer with respect to that direction are $R=8.5\times 10^{11}$ cm and $6.2\times 10^{11}$ cm, and $\arccos (\vec{e}_*\!\cdot \!\vec{e}_{\rm obs})=54\degr$ and $40\degr$ at $\theta=0.9$ and 2.0, respectively. 
} \label{taugg}
\end{figure}

As shown by \citet{cerutti11}, e$^\pm$ pair production on accretion blackbody disc photons in the \g-ray emission region is negligible. This process would absorb only \g-rays emitted from a vicinity of the accretion disc, at distances $\la 10^{8}$ cm for the parameters adopted here, or, at $\la 10^{10}$ cm if the disc emission get fully isotropized by the stellar wind (but see \citealt{sb11}). The present model does not predict \g-ray emission in these regions. At higher energies, \g-rays at $\ga 10$ GeV are above the threshold for pair production on stellar photons. The degree of attenuation strongly depends on the assumed inclination (which, in turn, is related to the fitted value of $H$ since the observed modulation depth close to unity requires that the star, the electron cloud and the observer are aligned at some phase.) For the parameters used here, the maximum optical depth, $\tau_{\rm \gamma\gamma}$, to this process is moderate, as shown in Fig.\ \ref{taugg}, which has been calculated using the method of \citet{dubus06} (taking into account the finite size of the star). However, an inclination $>30\degr$ would also yield a lower value of $H$, with both changes significantly increasing $\tau_{\rm \gamma\gamma}$.

\subsection{Magnetic field}
\label{bfield} 

We then consider the magnetic field in the \g-ray emitting region. We assume that the magnetic energy flux in the downstream region is a fraction, $\eta_B< 1$, of the dissipated power. Including both the jet and counterjet, the magnetic field in the jet comoving frame is given by,
\begin{equation}
{B^2\over 4} \beta_{\rm j}\Gamma_{\rm j}^2 c \Theta_{\rm j}^2 H^2=\eta_B \eta_{\rm diss} P_{\rm j,0}={\eta_B\over\eta_{\rm rad}\eta_{\rm e}}P_{\rm rad},
\label{bj0}
\end{equation}
where $\Theta_{\rm j}$ is the jet opening angle in the dissipation region, $\eta_{\rm rad}\simeq 1$, and the magnetic energy flux in the second equality is expressed in terms of the quantity closest to the observations, i.e., $P_{\rm rad}$. 

For $P_{\rm rad}$ estimated in Section \ref{j_power}, $B\simeq 120 (\eta_B/\eta_{\rm e})^{1/2}\Theta_{\rm j}^{-1}$ G, and $B^2/8\upi\simeq 560 (\eta_B/\eta_{\rm e})\Theta_{\rm j}^{-2}$ erg cm$^{-3}$. In comparison, the blackbody energy density within the electron cloud is $U_{\rm rad}\simeq 3.4\times 10^3 (T/10^5\,{\rm K})^4$ erg cm$^{-3}$, see equation (\ref{gdot1}). The magnetic field is constrained by the contribution of the synchrotron component to the broad-band spectra, which results in an upper limit on $\eta_B$, discussed below. 

The opening angle of the jet of Cyg X-3 on a 10 mas scale [$(d/7\,{\rm kpc})10^{15}$ cm] based on 2001 radio-outburst data of \citet{mj04} has been estimated as $\Theta_{\rm j}=5.0\pm 0.5\degr$ by \citet*{mj06}. We note that \citet{mj04} have fitted the jet curvature they observed as due to a jet precession with the precession angle of $2.4\degr$, which precession may smear the observed image, with the actual opening angle possibly being $<5\degr$. 

We approximate here the synchrotron spectrum using a delta-function approximation, in which the energy of a synchrotron photon averaged over the pitch angle, $\alpha$, is given by,
\begin{equation}
\epsilon= a {B\over B_{\rm cr}}\gamma^2, \quad a\simeq 1,\quad
B_{\rm cr}={2\upi m_{\rm e}^2 c^3\over e h},
\label{av_e}
\end{equation}
where $B_{\rm cr}$ is the critical magnetic field. This formula follows from the correspondence between the synchrotron and Compton processes \citep{bg70}, in which synchrotron emission is considered to be Compton scattering of virtual photons at the dimensionless energy of $B/B_{\rm cr}$. We then require that the power of the synchrotron emission in this approximation equals the actual synchrotron power, which then yields,
\begin{equation}
{\epsilon{\rm d} \dot n_{\rm S}\over {\rm d}\epsilon{\rm d}\Omega}\simeq
{\sigma_{\rm T} B_{\rm cr}^2 (B/B_{\rm cr})^{1/2} \epsilon^{1/2} \over 48\upi^2 m_{\rm e} c } N\left(\sqrt{\epsilon B_{\rm cr}\over B}\right)\,\,{\rm s}^{-1}.
\label{synspecdelta}
\end{equation}
Since $N(\gamma)$ gives the volume-integrated electron distribution, this formula gives the synchrotron emission from the entire source, analogously to the treatment in Section \ref{compton}. We have found that this formula provides a good approximation to the actual synchrotron spectrum. For a power-law spectrum, $N(\gamma)=K\gamma^{-p}$, with a value of $p>1/3$, the spectrum averaged over the pitch angle is (cf.\ \citealt*{jos74}),
\begin{eqnarray}
\lefteqn{
{\epsilon{\rm d} \dot n_{\rm S}\over {\rm d}\epsilon{\rm d}\Omega}\simeq  
C_1 {\sigma_{\rm T} c K B_{\rm cr}^2\over 48\upi^2 m_{\rm e} c} \left(B\over B_{\rm cr}\right)^{p+1\over 2}\epsilon^{-{p-1\over 2}}\,\,{\rm s}^{-1},\label{synspecpl}}\\
\lefteqn{
C_1 = {3^{p+4\over 2} \Gamma\left(3p-1\over 12\right) \Gamma\left(3p+19\over 12\right) \Gamma\left(p+1\over 4\right) \over 2^5\upi^{1\over 2}\Gamma\left(p+7\over 4\right)}\label{c1}},
\end{eqnarray}
where $\Gamma$ is the gamma function. In the delta-function approximation, equation (\ref{synspecdelta}), $C_1$ is set to 1. This approximation is fully accurate for $p=3$, for which $C_1(p)=1$. For $p=2$ and 4, $C_1(p)=1.14$, 1.20, respectively. For a broken electron power law, the break energy also appears close to the actual one. We note that $a\simeq 1$, which provides a phenomenological best fit to accurate results, does not correspond to the average synchrotron emission of a single electron, for which $a=2^2 3^{-1/2} 5^{-1} \sin\alpha$.

We need to take into account self-absorption of the synchrotron spectra. The synchrotron self-absorption coefficient in the jet frame for an electron power-law distribution averaged over the pitch angle can be expressed as (cf.\ \citealt{jos74}),
\begin{eqnarray}
\lefteqn{
\alpha_{\rm S}= C_2 {\upi \sigma_{\rm T} \over  2\alpha_{\rm f}}  {K\over V}\left(B\over B_{\rm cr}\right)^{p+2\over 2}\epsilon^{-{p+4\over 2}}\simeq {C_2 \upi \sigma_{\rm T} \over  2\alpha_{\rm f}V} {B_{\rm cr}\over B}\gamma^{-4} N\left(\gamma\right)
\,\,{\rm cm}^{-1},\label{alphas}}\\
\lefteqn{
C_2={3^{p+3\over 2} \Gamma\left(3p+2\over 12\right) \Gamma\left(3p+22\over 12\right) \Gamma\left(p+6\over 4\right)\over 2^4 \upi^{1/2}\Gamma\left(p+8\over 4\right)}},
\end{eqnarray}
where $\alpha_{\rm f}$ is the fine-structure constant, $V$ is the source volume in the jet or counterjet frame, $C_2\simeq 1$ for $p=3$. The second formula in equation (\ref{alphas}) gives $\alpha_{\rm S}$ in the monochromatic approximation, with $\gamma(\epsilon)$ given by equation (\ref{av_e}). This also gives $\alpha_{\rm S}$ corresponding to emission by electrons with a given $\gamma$. The volume depends on $\Delta H$, the length of the emission region along the jet in the observer's frame, and on the jet radius, $\Theta_{\rm j} H$. The value of $\Delta H$ is relatively uncertain; it is $\ga \beta_{\rm j}\Gamma_{\rm j}c \langle t_{\rm rad}\rangle\simeq 0.24 H$ based on the cooling argument (Section \ref{parameters}), and $\Delta H\ll H$ to account for the depth of the orbital modulation close to unity (DCH10). Given these constraints, we adopt 
\begin{equation}
\Delta H=\beta_{\rm j}\Gamma_{\rm j}c \langle t_{\rm rad}\rangle; \quad V\simeq \upi\Theta_{\rm j}^2 H^2 \Delta H\Gamma_{\rm j}.
\label{deltah}
\end{equation}
For our adopted parameters, $V\simeq 2.8\times 10^{33}$ cm$^3$. For $B$ given by equation (\ref{bj0}), we can determine the turnover energy (in the jet frame), at which the optical depth through the jet spine in the observer's direction, $\tau_{\rm S}(\epsilon_{\rm t})=2\alpha_{\rm S}\Theta_{\rm j} H/{\cal D}_{\rm j}\sin i=1$. We find this takes place between the radio and IR ranges, around $\sim 0.007(0.5\eta_B/10^{-3}\eta_{\rm e})^{3/10}$ eV, and in the part of the spectrum emitted by electrons dominated by adiabatic losses, $N(\gamma)\propto \gamma^{-1}$ of equation (\ref{ngamma1}). Below $\epsilon_{\rm t}$, $\dot n_{\rm S}\propto\epsilon^{3/2}$. 

The resulting spectra, from both the jet and the counterjet, and taking into account the relativistic transformation of equation (\ref{ef}), are shown in Figs.\ \ref{comparison}(a--b) for $\eta_B=10^{-3}$. The synchrotron spectrum has the shape similar to that of the Compton one, and, for the best-fit parameters, its peak, from electrons with $\gamma_1$, is at $\epsilon_{\rm S}\simeq 16 (\eta_B/\eta_{\rm e})^{1/2} (\gamma_1/10^3)^2$ eV.  We see that IR measurements simultaneous with those in \g-rays would provide constraints and/or a measurement of the jet magnetization, $\eta_B$. If the shown IR measurements during radio flares are representative for \g-ray active periods, the jet is relatively weakly magnetized, as an increase of $B$ would increase the synchrotron flux at the peak $\propto B^{(p+1)/2}$. We note that $\eta_B\sim 10^{-3}$ is consistent with the theoretical estimates for magnetized shocks of \citet{ml99}, \citet{medvedev06} and \citet{ss11}. At the above parameters and $\gamma_1= 10^3$, $B\simeq 60$ G. As a consequence of $\Delta H$ derived from electron cooling, the ratio of the magnetic field energy density to that in the electrons equals to $\eta_B/\eta_{\rm e}$, i.e., $(B^2/8\upi)/(E_{\rm e}/V)=\eta_B/\eta_{\rm e}$. The field strength is thus much below equipartition.

We also need to consider the synchrotron self-Compton process. The ratio of the energy density in the synchrotron photons, $U_{\rm S}$, to that in the magnetic field is,
\begin{equation}
{U_{\rm S}\over B^2/8\upi}\simeq {4\over 3\upi}{p-2\over p-3} {K\gamma_1^{3-p}\sigma_{\rm T}\over \Theta_{\rm j} H \Delta H\Gamma_{\rm j}},
\label{uratio}
\end{equation}
where $\Delta H$ is estimated as above, and the ratio is $\simeq 4$ for $p=4$, $\gamma_1=10^3$ and our adopted parameters. Thus, the self-Compton process is important, though Comptonization of blackbody radiation still dominates the electron losses, $(U_{\rm S}+ B^2/8\upi)/U_{\rm rad}\simeq 0.2$ [which reduces $\gamma_{\rm b}$, equation (\ref{gb2}), to $\simeq 10^2$] at $\eta_B= 10^{-3}$. Note that since the electron distribution is determined by the observed \g-ray spectrum (which is due to blackbody scattering), the above ratio is independent of $B$. 

We assume that the synchrotron emission is isotropic in the jet frame. The Compton process is here mostly in the Thomson limit, and we treat it using a delta-function approximation, $\epsilon=a' \gamma^2 \epsilon_0$, where $\epsilon_0$ and $\epsilon$ are the seed and scattered photon energy, respectively, and $a'=1$. This yields,
\begin{equation}
{\epsilon{\rm d} \dot n_{\rm SC}\over {\rm d}\epsilon{\rm d}\Omega}\simeq
{\sigma_{\rm T}c \epsilon^{1/2}\over 8\upi}\int^{\min(1/\epsilon,\epsilon)}_{\epsilon/\gamma_2^2} {n_{\rm S}(\epsilon_0)\over \epsilon_0^{1/2}} N\left(\sqrt{\epsilon \over \epsilon_0}\right){\rm d}\epsilon_0\,\,{\rm s}^{-1},
\label{synsc}
\end{equation}
where $n_{\rm S}$ is the density of the synchrotron photons,
\begin{equation}
n_{\rm S}(\epsilon_0)\simeq 4\upi {{\rm d} \dot n_{\rm S}\over {\rm d}\epsilon_0{\rm d}\Omega}{\Theta_{\rm j} H\over c V}\simeq {{\rm d} \dot n_{\rm S}\over {\rm d}\epsilon_0{\rm d}\Omega} {4\over c\Theta_{\rm j} H \Delta H\Gamma_{\rm j}}\,\,{\rm cm}^{-3},
\label{sdensity}
\end{equation}
and the integration limits in equation (\ref{synsc}) account for the Thomson limit (assuring that $\epsilon<\gamma$) and $\gamma_2\leq \gamma< 1$. We have found that using $a'=1$ reproduces better the exact Thomson-limit results for power law electrons (see Appendix \ref{isotropic}) for $2\leq p\leq 4$ than $a'=4/3$, corresponding to the average scattered energy. The resulting spectra, from both the jet and the counterjet, and taking into account the relativistic transformation of equation (\ref{ef}), are shown in Figs.\ \ref{comparison}(a--b). We see that the self-Compton component may contribute to both the X-ray high-energy tail and the \g-ray spectrum above 0.1 GeV. The relative strength of this component is constrained by the orbital modulation. Intrinsically, the self-Compton component is not orbitally modulated, which implies its contribution to the GeV range (with strong modulation) is weak. We find that the contribution of the self-Compton component to the range $>0.1$ TeV (not shown in Figs.\ \ref{comparison}a--b) is below the extrapolation of the \fermi/LAT power law. In the hard X-rays, we also see a relatively strong orbital modulation due to wind absorption \citep{z12}, which appears to imply that the tail is not mainly due to this process. These constraints on the relative amplitude of the self-Compton component also give an upper limit on $B$, as its increase would amplify both the synchrotron and self-Compton components by $\sim B^{(p+1)/2}$. Thus, we obtain $B\la 10^2$ G within the \g-ray emitting source.

We also calculate the radial Thomson optical depth of the electrons, $\tau_{\rm T}=N_{\rm e}/(2\upi\Theta_{\rm j} H \Delta H\Gamma_{\rm j})$, which is $\simeq 7\times 10^{-5}$ for our adopted parameters. 

\section{Discussion}
\label{discussion}

\subsection{Electron acceleration}
\label{acceleration}

It remains unclear what mechanism is responsible for acceleration of electrons producing high energy gamma-rays in jets. It is often considered to be diffusive shock acceleration (DSA), which involves the first order Fermi process (e.g., \citealt{bo78,be87}). Initial studies of DSA scenarios were focused on explaining the origin of cosmic rays. The process was intensively explored by Monte Carlo simulations to take into account different shock parameters and magnetic field structures (e.g. \citealt{no04} and references therein). Such simulations fully confirmed the ability of DSA to produce ultrarelativistic cosmic rays, but acceleration of electrons (and positrons) was achieved only after assuming that they were preheated up to energies corresponding with the momentum of thermal, shocked ions. Recent results obtained using particle in cell (PIC) simulations have shown that in collisionless relativistic electron-ion shocks a quasi-Maxwellian distribution of electrons is produced, with the average energy of $\bar \gamma_{\rm inj} \la (m_{\rm i}/m_{\rm e}) \bar \gamma_{\rm i}$ (\citealt{ss11}, see also \citealt{s08}), where $\bar \gamma_{\rm i}$ is the average Lorentz factor of ions. A quasi-Maxwellian distribution of preheated electrons with the temperature close to that of the ions is also found in PIC simulations of non-relativistic shocks \citep{rs11}. This provides the required preheating, solving the above long-standing problem.

There are still no available results of PIC simulations of mildly relativistic shocks. However, given the results mentioned above for relativistic and non-relativistic shocks, we can assume that a quasi-Maxwellian distribution of efficiently preheated electrons is produced also in mildly relativistic shocks, which is likely to be the case in the Cyg X-3 jet. Noting that the relative contribution of the low-energy tail of a Maxwellian distribution to observed electromagnetic spectra is small, we have approximated the electron injection spectrum to have a cut-off at an energy of $\gamma_1 \sim \bar \gamma_{\rm inj}$. Then, our finding of $\gamma_1\sim 300$--$10^3$ is consistent with the above results, with this value being related to the $m_{\rm i}/m_{\rm e}$ mass ratio. 

We note that low-energy breaks at $\gamma_1\gg 1$ are common in jets of AGNs, where they are also often attributed to the ion mass (e.g., \citealt{stawarz07}). \citet{ghisellini10,ghisellini11} find that the electron distribution of blazars observed by \fermi\/ commonly show a relatively steep injection above $\gamma_1\sim 10^2$--$10^3$ (denoted in their papers by $\gamma_{\rm b}$), and a hard injection below it, for which the steady-state distribution is approximately compatible with that of our equation (\ref{ngamma1}). The low-energy cut-offs/breaks at $\gamma_1 \sim 10^2$--$10^3$ are observed not only in blazars but also in spectra of hot spots in radio-lobes \citep{blundell06,stawarz07,godfrey09} and used to argue for the presence of protons. Finally, we note that since the mass per ion in Cyg X-3 (a helium system) is $\sim$3 times higher than for the cosmic abundances, the value of $\gamma_1$ in it may be correspondingly higher than in comparable systems with abundances dominated by hydrogen. 

\subsection{Caveats}
\label{caveats}

DCH10 obtained some ranges of the allowed parameters, but, given the complexity of the problem, we have just used the best-fit values (adjusted for the case of a steady jet) in this study. Furthermore, the best-fit parameters of DCH10 may be modified if the minimum of the X-ray folded light curve does not exactly correspond to the superior conjunction. This may happen if the wind is not symmetric with respect to the conjunctions, e.g., due to the wind lagging the binary rotation, and/or a formation of a Compton cloud around the compact object, as in the model of \citet{zmg10}. 

There is then a considerable uncertainty regarding the binary parameters of Cyg X-3. We have adopted the binary parameters for the black-hole case used by DCH10, in particular the mass of the WR star of $50\msun$, which they assumed following \citet{sz08}. We note that such high mass appears inconsistent with the mass-loss rate of $\sim 10^{-5}\msun/$yr, estimated for Cyg X-3 (e.g., \citealt{sz08} and references therein). The mass vs.\ mass-loss rate relationship in WR stars is $\dot M\sim 10^{-7}(M_*/\msun)^m \msun/$yr, with $m\simeq 2$--2.5 \citep{langer89,sm92}. This implies $M_*\sim 10\msun$, which also agrees with the results of \citet{lommen05} (though \citealt*{hsf00} favour a higher mass in the black-hole case). However, the value of $M_*$ affects only relatively slightly the orbital separation, $\propto (M_*+M)^{1/3}$. 

The stellar luminosity adopted by DCH10, $\simeq 1.8\times 10^{39}$ erg s$^{-1}$, is close (within a factor of 2) to that predicted by the WR mass-luminosity relation \citep{sm92}. The chosen temperature, $10^5$ K, corresponds to the hydrostatic stellar surface rather than the photosphere of an isolated star, with the effective temperature of the photosphere a few times lower \citep{sm92}. However, the X-ray source in Cyg X-3 strongly ionizes the wind on the side of the jet, and thus the jet is likely to be exposed to radiation at the temperature close to the core one. On the other hand, Compton scattering of the stellar radiation by the wind will be substantial, which will increase the apparent size of the diluted blackbody source.

In this study, we have used the Klein-Nishina cross section for calculating spectra, but still assumed the steady-state electron distribution is a power law. We note that such an approach is not fully self-consistent if the energy losses of the electrons by Compton scattering dominate over the synchrotron and adiabatic ones \citep{zk93,moderski05}. If the electrons are accelerated at a power-law rate, the electron energy losses reduced in the Klein-Nishina regime cause the steady-state electron distribution to be no longer a power law. In fact, the reduced energy losses are largely compensated by the reduced Compton scattering emission, and the final spectrum is approximately a power law with the same index and normalization as that in the Thomson regime \citep{zk93,moderski05}. Note that this effect would need to be taken into account in calculating constraints on the maximum accelerated energy, $\gamma_2$.

Finally, we have found that a model with a low value of the low-energy cut-off in the electron distribution, $\gamma_1<\gamma_{\rm b}$, and the break energy due to energy losses at $\gamma_{\rm b}\simeq 10^3$ can also explain the \g-ray spectrum of Cyg X-3. Although we cannot rule out this model, we consider it unlikely given our estimate of $\gamma_{\rm b}\simeq 10^2$. 

\subsection{The uniqueness of $\mathbf{\gamma}$-ray emission of Cyg X-3}
\label{unique}

We briefly address the question why Cyg X-3 is, so far, the only accreting X-ray binary with confirmed high-energy \g-ray emission. We note that although a number of other X-ray binaries, e.g., LS I +61{\degr}303 or LS 5039 emit high-energy \g-rays, that emission is, most likely, due to collision of their pulsar winds with stellar winds of their high-mass companions rather than due to accretion (e.g., \citealt*{dubus06b,nc07,znc10}). 

DCH10 noted that since Cyg X-3 has both a very high wind mass loss rate and a very small separation, it may be unique in forming a reconfinement shock in its inner jet. We note that, for the adopted Cyg X-3 parameters, this requires a relatively large initial jet opening angle for a reconfinement shock to occur, $\Theta_{\rm j}\ga 30\degr$. This follows from equation (7) of DCH10, which implies $H>R$ for a smaller $\Theta_{\rm j}$, while $H< R$ from the geometry of the system. Such a large initial opening angle may be formed in the jet formation mechanism utilizing disc magnetic field \citep{bp82}. We note that a similar initial wide opening jet angle is seen in the radio galaxy M87 \citep*{biretta02}.

A related unique feature of Cyg X-3 is the very large luminosity of its companion ($L_*\sim 10^{39}$ erg s$^{-1}$) accompanying its very small separation. This results in a very high blackbody flux irradiating the \g-ray emitting region, which then yields a strong Compton-scattering flux. If the orbital separation were $\ga 10$ times higher (as in Cyg X-1) or the stellar luminosity were much lower (as in low-mass X-ray binaries), synchrotron and self-Compton emission would dominate instead of blackbody up-scattering (see Section \ref{bfield}). The presence of such synchrotron component may be searched for in those systems. The relative strength of the associated self-Compton component in jets of those objects during dissipation events remains unknown (as it depends on the unknown jet parameters); we note it might produce observable \g-ray emission.

\subsection{Relationship to radio emission}
\label{radio}

After the relativistic electrons lose their energy at $H\sim 10^{12}$ cm, the jet continues to propagate for a large distance until another dissipation region forms. Since the stellar emission is very weak at that point, the main energy losses are synchrotron. The characteristic size of the sources of the resulting flaring radio emission is $\sim 10^{15}$ cm, and the variability is on a day time scale (e.g., \citealt{mj04}). In Fig.\ \ref{r_x_gamma}, we see that the average radio luminosity during the active periods is much lower than the \g-ray luminosity. Thus, the observed radio emission is energetically allowed to be emitted by the jet downstream the \g-ray dissipation region, with its power reduced by about an order of magnitude (Section \ref{j_power}) with respect to the upstream jet.

If the jet experiences a dissipation episode of the kind studied here, its velocity in the radio-emitting region should be similar to that in the dissipation region (estimated from the orbital modulation), $\beta_{\rm j}\sim 0.5$. Indeed, such a velocity has been estimated from the proper motion by \citet{mj06}, and a similar $\beta_{\rm j}\simeq 0.6$ was estimated from fitting a precession model to a radio image obtained during a flaring state \citep{mj04}. Similar estimates have been obtained from a number of other radio observations of Cyg X-3, e.g., $\beta_{\rm j}\simeq 0.5$ inferred by \citet*{marti01}, except for \citet{m01}, who estimated $\beta_{\rm j}\ga 0.8$. We note that our estimate of the jet velocity {\it before\/} the dissipation region is compatible with that, $\beta_{\rm j,0}\sim 0.9$. It is then possible that the measurement of \citet{m01} was done during a radio flaring episode during which the dissipation region producing \g-rays was not formed, and the jet propagated to large distances with $\beta_{\rm j,0}$. Occurrence of radio flaring episodes without formation of a prior dissipation region at scales $\ll 10^{15}$ cm can also explain a radio flare occurring before a \g-ray flare \citep{williams11}.

Recently, mm radio flares lasting a fraction of a day and occurring on intermediate size scales, $\sim 10^{13}$ cm, have been discovered \citep{tsuboi10,tsuboi12}. This size scale is derived from the flare rise time scales of several minutes. The second flare of \citet{tsuboi12}, observed at 43 GHz and 86 GHz, took place on MJD 54972.9 during a period quiescent in both radio emission at 15 GHz and in \g-rays (FLC09). Thus, this is an example of a radio flare without a prior jet energy dissipation on the orbital size scale. The electron power required was found $>3\times (D/7\,{\rm kpc})^2 10^{37}$ erg s$^{-1}$ (neglecting relativistic corrections), similar to our estimate of the electron power. The magnetic field was estimated at $\sim 10$ G (assuming equipartition), an order of magnitude below our upper limit for the \g-ray emitting region. 

\section{Conclusions}
\label{conclusions}

We have obtained the average X-ray spectrum and the radio flux emitted during the \g-ray active epochs (Section \ref{data}). We have then calculated spectra from Compton scattering of a photon beam into a given direction by isotropic relativistic electrons with a power-law distribution with a low-energy cut-off. Simple analytical formulae have been obtained both using the Klein-Nishina cross section (Section \ref{compton}) and in the Thomson limit (Appendix \ref{thomson}). 

We have applied our results to scattering of stellar blackbody radiation by relativistic electrons in the jet of Cyg X-3 (Section \ref{observed}), using the model of DCH10, fitted to the observed modulation of \g-rays (FLC09). We have found a low-energy break at $\gamma_1\sim 300$--$10^3$ in the distribution of the accelerated electrons is required by the observational data in order not to overproduce the observed X-ray emission. We find Compton cooling to be efficient, which implies the power-law index of the acceleration process of $\simeq 2.5$--3, rather typical to astrophysical acceleration sites. The low-energy electron break found by us is in agreement with recent shock acceleration models, in which it is related to the ion/electron mass ratio. Also, the obtained value of the break Lorentz factor is similar to those typically found in AGN jets (see Section \ref{acceleration}).

We have calculated the jet kinetic power to be $\sim 10^{38}$ erg s$^{-1}$ assuming equipartition between the energy supplied to the electrons and ions, which represents a firm lower limit. The estimated power is comparable to the radiative power of this source. Assuming this equipartition, the bulk Lorentz factor of the jet before the dissipation region is $\sim 2.5$. Most of the power supplied to the electrons is radiated (Sections \ref{j_power}--\ref{parameters}). 

We have found that the magnetic field strength is constrained to be below the equipartition with the electron energy density by a factor of $\la$ a few times $10^{-3}$. At the upper limit ($B\sim 10^2$ G), the synchrotron emission from the \g-ray emitting region still gives rise to a relatively strong IR flux, which measurement simultaneous with that of \g-rays would provide an estimate of the magnetic field in this part of the jet (Section \ref{bfield}). The predicted synchrotron flux is at the level at the IR flux measured during past radio flares of Cyg X-3.

\section*{ACKNOWLEDGMENTS}

We thank G. Skinner, A. Szostek and S. Corbel for the BAT, \sax, and radio data, respectively, A. Frankowski and P. Lubi{\'n}ski for help with the analysis of the BAT data, and R. Misra and R. Moderski for valuable discussions. We thank the referee for stimulating comments. This research has been supported in part by the Polish NCN grants N N203 581240, N N203 404939, and 362/1/N-INTEGRAL/2008/09/0. FY has been supported by the NSF of China (grants 10821302, 10825314, 11133005).

\appendix

\section{The Thomson limit}
\label{thomson}

In the Thomson limit, the flux per electron and per solid angle becomes,
\begin{eqnarray}
\lefteqn{
{\epsilon{\rm d} \dot n(\epsilon_0,\gamma)\over {\rm d}\epsilon{\rm d}\Omega}= 
\cases{\displaystyle{{3\sigma_{\rm T}\over 8\upi} \dot n_0 x r \left[\left(1-r\right)^2+r^2\right]} , &$2\epsilon_0 x \gamma<1$;\cr
0, &$2\epsilon_0 x \gamma\geq 1$,\cr}\label{rate0}
}\\
\lefteqn{
r= {\epsilon\over 2\epsilon_0\gamma^2 x},}
\end{eqnarray}
where, in order to approximately constrain the resulting spectrum to the energy range satisfying the Thomson limit, we have imposed a sharp cut-off at the range boundary. This, in particular, assures $\epsilon<\gamma$. Then, the condition of $r\leq 1$ yield the electron-integrated rate given by equation (\ref{integration}) but with the lower and upper limits of
\begin{equation}
\max[\gamma_1, (\epsilon/2 x\epsilon_0)^{1/2}],\quad \max\{\min[\gamma_2, 1/(2 x\epsilon_0)],\gamma_1\},
\label{limits}
\end{equation}
respectively. These limits, the constraint of $\epsilon\leq 2\epsilon_0 x\gamma_2^2$, and a Thomson limit condition of $\epsilon\leq 1/(2 x\epsilon_0)$ yield,
\begin{eqnarray}
\lefteqn{
{\epsilon {\rm d} \dot n(\epsilon_0)\over {\rm d}\epsilon{\rm d}\Omega}= {3\sigma_{\rm T}\over 8\upi} \dot n_0 K \times\nonumber}\\
\lefteqn{
\cases{
f_1-f_3, &$\displaystyle{\min\left({\epsilon\over \gamma_2^2},{1\over \epsilon}\right)< 2 x\epsilon_0< \min\left({\epsilon\over \gamma_1^2},{1\over \gamma_2},{1\over \epsilon}\right)}$;\cr
f_1-f_4, &$\displaystyle{\min\left({\epsilon\over \gamma_1^2},{1\over \gamma_2},{1\over \epsilon}\right) < 2 x\epsilon_0< \min\left({\epsilon\over \gamma_1^2},{1\over \epsilon}\right) }$,\cr
f_2-f_3, &$\displaystyle{\min\left({\epsilon\over \gamma_1^2},{1\over \epsilon}\right)< 2 x\epsilon_0< \min\left[{1\over \epsilon},\max\left({\epsilon\over \gamma_1^2},{1\over \gamma_2}\right)\right]}$,\cr
f_2-f_4, &$\displaystyle{ \min\left[{1\over \epsilon},\max\left({\epsilon\over \gamma_1^2},{1\over \gamma_2}\right)\right]< 2 x\epsilon_0<{1\over \epsilon}}$,\cr
0, &$\displaystyle{2 x\epsilon_0<{\epsilon\over \gamma_2^2}\,\, {\rm or}\,\, 2 x\epsilon_0>{1\over \epsilon}\,\, {\rm or}\,\,2 x\epsilon_0>{1\over \gamma_1}}$,\cr}
\label{power}}\\
\lefteqn{
f_1=\left(\epsilon\over 2\epsilon_0\right)^{{1-p}\over 2}x^{{1+p}\over 2}
{11+4p+p^2\over (1+p)(3+p)(5+p)},}\\
\lefteqn{f_2=
{\epsilon\gamma_1^{-1-p}\over 2\epsilon_0(1+p)}-{\epsilon^2\gamma_1^{-3-p}\over 2\epsilon_0^2 x(3+p)}+{\epsilon^3\gamma_1^{-5-p}\over 4\epsilon_0^3 x^2(5+p)},}\\
\lefteqn{f_3={\epsilon \gamma_2^{-1-p}\over 2\epsilon_0(1+p)}-
{\epsilon^2 \gamma_2^{-3-p}\over 2\epsilon_0^2 x(3+p)}
+{\epsilon^3 \gamma_2^{-5-p}\over 4\epsilon_0^3 x^2(5+p)},}\\
\lefteqn{f_4={\epsilon (2\epsilon_0 x)^{1+p}\over 2\epsilon_0(1+p)}-
{\epsilon^2 (2\epsilon_0 x)^{2+p}\over \epsilon_0 (3+p)}
+{\epsilon^3 (2\epsilon_0 x)^{3+p}\over \epsilon_0(5+p)}.}
\end{eqnarray}
When $\epsilon<\gamma_1^2/\gamma_2$, there is no range for $f_1-f_4$, and when $\epsilon>\gamma_1^2/\gamma_2$, there is no range for $f_2-f_3$. When $\gamma_2\rightarrow \infty$, only the $f_1-f_4$ and $f_2-f_4$ ranges exist. When $\epsilon\geq \gamma_2$, the spectrum is null. The above formulae should be applied for $\epsilon\gg \epsilon_0$ only. The emitted spectrum in the part dominated by $f_1$ is a power law with an energy index of $\alpha=(p-1)/2$. 

Figs.\ \ref{mono} and \ref{beam} shows compares the spectra obtained using equations (\ref{power}) and (\ref{int_rate}) for blackbody irradiation, respectively, with the corresponding ones obtained using the Klein-Nishina cross section. When $\epsilon/ (2\gamma_2^2 x)\ll\langle \epsilon_0 \rangle\ll \min[\epsilon/ (2\gamma_1^2 x),1/(2 x\epsilon)]$, where $\langle\epsilon_0\rangle$ is the characteristic energy of the seed photons, the integral over $f_1$ in equation (\ref{int_rate}) dominates, and we can set its limits from zero to infinity. 

In the case of diluted blackbody seed photons, equation (\ref{bb}), and for $\epsilon/ (2\gamma_2^2 x)\ll kT/(m_{\rm e}c^2 {\cal D}_*) \ll \min[\epsilon/ (2\gamma_1^2 x),1/(2 x\epsilon)]$, the scattered flux in a given direction per dimensional energy in the jet frame becomes,
\begin{eqnarray}
\lefteqn{
{\epsilon {\rm d} \dot n\over {\rm d}\epsilon {\rm d}\Omega}=2^{p-9\over 2}3 \sigma_{\rm T} c^{-2} h^{-3}  K (\epsilon m_{\rm e}c^2)^{{1-p}\over 2} (kT)^{5+p\over 2}\left(x\over {\cal D}_*\right)^{{1+p}\over 2} \left(R_*\over R\right)^2 \times  \nonumber}\\
\lefteqn{\qquad \quad {11+4p+p^2\over 5+p}  \Gamma\left(1+p\over 2\right)\zeta\left(5+p\over 2\right),\label{bb_rate}}
\end{eqnarray}
where $\zeta$ is the Riemann function. This flux was obtained by DCH10, see their equations (1), (3). [Their formula is in the observer's frame assuming the ${\cal D}_{\rm j}^3$ transformation, see Section \ref{observed}, it is $4\upi$ times larger than ours, which results from their definition of $K$ as corresponding to ${\rm d} N/{\rm d}\gamma{\rm d}\Omega$ rather than ${\rm d} N/{\rm d}\gamma$, and the power of 2 in their equation (3) is misprinted as $p+5/2$, while is should be $(p+5)/2$.] 

\section{Isotropic scattering}
\label{isotropic}

We note that integrating spectra over the scattering angle gives either the spectrum for a photon beam integrated over all directions of the scattered photon or the spectrum from scattering on an isotropic seed photon distribution. Also, the case of isotropic seed photons and isotropic electrons is equivalent to an electron beam when scattered photons are integrated over all directions. 

Therefore, integrating equation (\ref{rate0}) over $\cos\vartheta$ in the range corresponding to $r\leq 1$ and multiplying by $2\upi$ (corresponding to integration over the azimuth), we obtain the isotropic rate of equation (2.42) of \citet{bg70}. Integrating $f_i$ of equation (\ref{power}) over $x$ and multiplying by $2\upi$ gives power-law electron rates integrated over all directions, which also correspond to emission at any direction in the case of isotropic seed photons. In particular, assuming that we are far below both the maximum emitted $\epsilon$ by electrons with $\gamma_2$ and the boundary of the Thomson limit, we obtain,
\begin{eqnarray}
\lefteqn{
{\epsilon {\rm d} \dot n\over {\rm d}\epsilon}= {3\sigma_{\rm T}\over 8\upi} \dot n_0 K\times \cases{f_1^{\rm iso}(\epsilon,\epsilon_0,p), &$\epsilon\geq 4\epsilon_0 \gamma_1^2$;\cr
f_2^{\rm iso}(\epsilon,\epsilon_0,p,\gamma_1), &$\epsilon\leq 4\epsilon_0 \gamma_1^2$,\cr}\label{power_iso}}\\
\lefteqn{
f_1^{\rm iso}=2\upi \int_0^2 f_1 {\rm d}x=2^{3+p}\upi \left(\epsilon\over \epsilon_0\right)^{{1-p}\over 2}\!\!
{11+4p+p^2\over (1+p)(3+p)^2(5+p)}\label{power_iso1} }\\
\lefteqn{
f_2^{\rm iso}=2\upi \int_0^{\epsilon/(2\epsilon_0\gamma_1^2)} f_1 {\rm d}x +
2\upi \int_{\epsilon/(2\epsilon_0\gamma_1^2)}^2 f_2 {\rm d}x ={\upi\over 4\gamma_1^{5+p}}\times \label{power_iso2}}\\
\lefteqn{ \left[{8\epsilon \gamma_1^4\over \epsilon_0(1+p)}+{2\epsilon^2\gamma_1^2\over \epsilon_0^2(3+p)}\left(1+2\ln{\epsilon\over 4\epsilon_0\gamma_1^2}\right) -{8\epsilon^2\gamma_1^2\over \epsilon_0^2 (3+p)^2} - {\epsilon^3\over \epsilon_0^3(5+p)} 
\right] \nonumber}
\end{eqnarray}
These rates can also be obtained by integrating the isotropic rate of equation (2.42) of \citet{bg70} over the electron distribution (\ref{ngamma}). The rate of equation (\ref{power_iso1}) is given by equation (2.64) of \citet{bg70}. We can then integrate over a photon distribution, 
\begin{eqnarray}
\lefteqn{
{\epsilon {\rm d} \dot n\over {\rm d}\epsilon}= {3\sigma_{\rm T}\over 8\upi} K\left[ \int_0^{\epsilon/ (4\gamma_1^2)} \dot n_0(\epsilon_0) f_1^{\rm iso}(\epsilon,\epsilon_0,p) {\rm d}\epsilon_0\right.\nonumber}\\
\lefteqn{\qquad\qquad 
+\left.\int_{\epsilon/ (4\gamma_1^2)} \dot n_0(\epsilon_0) f_2^{\rm iso}(\epsilon,\epsilon_0,p,\gamma_1) {\rm d}\epsilon_0\right].}
\label{int_rate_iso}
\end{eqnarray}

Integrating the rate of equation (\ref{bb_rate}) over $x$ from 0 to 2 and multiplying by $2\upi$ gives the isotropic rate of equations (2.65--2.66) of \citet{bg70}), except that their rate is given for the seed photon density inside a blackbody field, with $n_0(\epsilon_0)$ in units of erg$^{-1}$ cm$^{-3}$, rather for the photon flux, as in our equation (\ref{bb}). That rate can also be obtained from equation (\ref{int_rate_iso}) for $\epsilon/ (4\gamma_1^2)\gg kT/{\cal D}_*$ (i.e., neglecting the integral over $f_2^{\rm iso}$). Integrating the Klein-Nishina rate of equation (\ref{pl_rate}) over the scattering angle leads to the isotropic Klein-Nishina rate of equations (25) and (A9) of \citet{aa81}.

\label{lastpage}
\end{document}